\documentclass{pepsart}

\usepackage[utf8]{inputenc} 
\usepackage{wasysym}
\usepackage{natbib}
\usepackage{chemfig}
\usepackage{lineno}
\usepackage{amssymb}
\usepackage{amsmath}
\usepackage{mathrsfs}
\usepackage{commath}
\usepackage{mathtools}
\usepackage{dutchcal}
\usepackage{stmaryrd}
\usepackage{pifont}
\usepackage{url}
\usepackage{graphicx}
\usepackage{subcaption}
\usepackage[noabbrev]{cleveref}
\usepackage{csquotes}
\usepackage{adjustbox}
\usepackage{textcomp}
\usepackage{makecell}

\usepackage{algorithm}
\usepackage{algpseudocode}

\algdef{SE}[DOWHILE]{Do}{doWhile}{\algorithmicdo}[1]{\algorithmicwhile\ #1}%
\algnewcommand\algorithmicforeach{\textbf{for each}}
\algdef{S}[FOR]{ForEach}[1]{\algorithmicforeach\ #1\ \algorithmicdo}

\usepackage{siunitx}

\usepackage{pgfplots}
\pgfplotsset{compat=1.16}
\usepackage{booktabs}
\usepgfplotslibrary{polar}
\usetikzlibrary{calc}

\DeclareMathOperator*{\argmin}{arg\,min}

\DeclareMathOperator*{\erf}{erf}

\renewcommand\vec{\boldsymbol}

\DeclareSIUnit\year{yr}

\def\cite{\citep}

\usepackage[acronym]{glossaries-prefix}
\makenoidxglossaries

\newacronym{FE}{FE}{Finite Element}
\newacronym{PDE}{PDE}{Partial Differential Equation}
\newacronym{UFL}{UFL}{Unified Form Language}
\newacronym{FFC}{FFC}{FEniCS Form Compiler}
\newacronym[prefixfirst={the\ }]{PETSc}{PETSc}{Portable, Extensible Toolkit for Scientific Computation}
\newacronym{CAD}{CAD}{Computer--Aided Design}
\newacronym{DoF}{DoF}{Degree of Freedom}
\newacronym[prefixfirst={a\ }, prefix={an\ }]{FGMRES}{FGMRES}{Flexible Generalized Minimal Residual}
\newacronym[prefixfirst={a\ }, prefix={a\ }]{GMRES}{GMRES}{Generalized Minimal Residual}
\newacronym{ILU}{iLU}{incomplete LU}
\newacronym{BSpline}{B-spline}{B\'{e}zier spline}

\begin{document}

\begin{frontmatter}

\begin{fmbox}
\dochead{Research}


\title{Thermal modeling of subduction zones with prescribed and evolving 2D and 3D slab geometries}


\author[
  addressref={aff1},
  email={nsime@carnegiescience.edu}
]{\inits{N}\fnm{Nathan} \snm{Sime}}
\author[
  addressref={aff1},
  email={cwilson@carnegiescience.edu}
]{\inits{C.R.}\fnm{Cian R.} \snm{Wilson}}
\author[
  addressref={aff1},
  corref={aff1},
  email={pvankeken@carnegiescience.edu}
]{\inits{P.E.}\fnm{Peter E.} \snm{{van~Keken}}}

\address[id=aff1]{
  \orgname{Earth and Planets Laboratory, Carnegie Institution for Science},
  \street{5241 Broad Branch Road NW},
  \city{Washington, DC},
  \postcode{20015},
  \cny{USA}
}
\address[id=aff2]{}


\begin{artnotes}

\end{artnotes}

\end{fmbox}


\begin{abstractbox}

\begin{abstract} 
The determination of the temperature in and above the slab in subduction zones,
using models where the top of the slab is precisely known,
is important to test hypotheses regarding the causes of arc volcanism and
intermediate-depth seismicity. While 2D and 3D models can predict the thermal
structure with high precision for fixed slab geometries, a number of regions
are characterized by relatively large geometrical changes. Examples include
the flat slab segments in South America that evolved from more steeply dipping
geometries to the present day flat slab geometry.
We devise, implement, and test a numerical approach
to model the thermal evolution of a subduction zone with prescribed changes in
slab geometry over time.
Our numerical model approximates the subduction zone geometry by
employing time dependent deformation of a B\'{e}zier spline which is used as
the slab interface in a finite element discretization of the Stokes and heat
equations. We implement the numerical model using the FEniCS open source
finite element suite and describe the means by which we compute approximations
of the subduction zone velocity, temperature, and pressure fields. We compute and
compare the 3D time evolving numerical model with its 2D analogy at 
cross-sections for slabs that evolve to 
the present-day structure of a flat segment of the subducting Nazca plate.
\end{abstract}




\end{abstractbox}
%

\end{frontmatter}

\nocite{settings}




\section{Introduction}
\label{sec:introduction}
\vspace{6pt}


\subsection{The importance of the thermal structure of flat slab segments}

Upon subduction the oceanic lithosphere warms and undergoes metamorphic 
phase changes which release fluids that can lead to arc volcanism and
intermediate-depth seismicity.
These regions have
significant potential for major natural hazards that include underthrusting
seismic events along the plate interface and explosive arc volcanism.
The geophysical and geochemical processes that cause such hazards are strongly
controlled by temperature 
\citep[for a recent review see][]{vanKekenWilson2023} 
and it is of great interest to a broad community
of Earth scientists to understand the thermal structure of subduction zones. 

Of particular interest to us is the intermediate-depth and deep seismicity 
that occurs at depths below the brittle-ductile transition and require mechanisms
other than brittle failure. 
Shear heating instabilities 
\cite{KelemenHirth2007},
dehydration embrittlement 
\cite{RaleighPaterson1965,Jung2004},
and hydration-related embrittlement 
\citep[e.g.,][]{Shiina2013,vanKeken2012,Shirey2021}
are three such mechanisms. The difference between the last two is that dehydration embrittlement would limit the seismicity
to the location of metamorphic dehydration whereas hydration-related embrittlement may occur wherever fluids 
that have been liberated by such dehydration reactions migrate. The wide distribution of these fluids that is predicted 
by fluid flow modeling \cite{wilson2014},
observed at locations of intermediate-depth seismicity 
\cite{Shiina2013,Shiina2017,Bloch2018},
and seen in petrological studies of exhumed
oceanic crust 
\cite{BeboutSPD2016}
provides strong support for the latter hypothesis. Thermal modeling suggests intermediate-depth
seismicity is limited to be above major dehydration reactions such as that of blueschist-out
or antigorite-out phase boundaries
\cite{vanKeken2012,Wei2017,Sippl2019}.
Further indications of petrological controls on the locations of seismicity are provided
by 
\citet{Abers2013}
who showed that the upper plane of seismicity in cold subduction zones tends to be limited to the oceanic
crust whereas the seismicity in warm subduction zones occurs in the mantle portion of the subducting slab. 
See \citet{vanKekenWilson2023}
for a broader discussion of the relationship between metamorphic dehydration reactions, fluids, and intermediate-depth seismicity.

The observational evidence, combined with modeling constraints strongly suggests fluids and intermediate-depth earthquakes
are related but how can this relationship be further constrained and quantified? 
\citet{Wagner2020} lays out an elegant motivation that 
a number of flat slab regions provide natural experiments to study this question. 
In a number of regions on Earth, such as below Southern Alaska \cite{Finzel2011},
Colombia \citep{Wagner2017}, 
Mexico, Peru, and Chile \cite{Manea2017},
subduction zones are characterized by flat slabs
where upon subduction the slab top stays flat after it reaches a certain depth below the continental lithosphere over
significant distances. 
Periods of flat slab subduction in the geological past have also been suggested to cause orogenic events and ore deposits
far from paleogeographically constrained plate boundaries.
These include the late Cretaceous to Paleocene Laramide orogeny \citep[see, e.g.,][]{Fan2014,Carrapa2019}
and the Mesozoic South China fold belt \cite{LiLi2007}.
It is generally understood that these slab segments
form by trench rollback with the continental lithosphere overriding the slab at
shallow depth causing effective flattening.

Specific modern-day flat slab regions that would allow us to further quantify the fluid-seismicity relationship 
include the present-day Pampean slab (beneath Chile and Argentina) and the
Peruvian flat slab segment. Both likely evolved from steeper subduction to near 
flat subduction due to the subduction of more buoyant thickened ridges such as
the Juan Fern\'andez and Nazca Ridges \cite{Gutscher1999, Antonijevic2015,
ContrerasReyes2019}. These regions are of particular interest as the
intermediate-depth seismicity varies significantly along-trench. This may be
caused by a variable hydration state of the incoming lithosphere and the thermal
evolution of the subduction crust and mantle
\citep[see][for observational evidence and the development of a testable hypothesis]{Wagner2020}.
These locations therefore suggest at least a qualitative correlation between fluids and seismicity.
In order to further test and quantify this correlation 
we need a good understanding of the thermal structure of the flat slab as it evolves.

Numerical modeling provides an important complement to observational studies
as it can predict the subduction zone thermal structure by computing approximations of partial 
differential equations' solutions that arise from the conservation principles of mass,
momentum, and thermal energy. Thermal models of flat slab segments have provided 
insights into the thermal evolution of the slab and overriding lithosphere but
have generally been predicted using 2D cross-sections \citep[e.g.,][]{English2003,Manea2011,Marot2014,Axen2018,Liu2022,Currie2022}. 
The geological evolution of the South Peruvian and Pampean slabs is influenced by strong temporal changes 
in 3D geometry over time. This makes reliable predictions of their thermal evolution
challenging as it needs to be approached with methods that can prescribe such 3D 
geometrical evolution in a paleogeographically consistent fashion.
A few 3D model simulations exist for flat slab reconstructions \cite{Schmid2002,Liu2008}
or their geodynamical evolution \cite{Jadamec2013,Haynie2017,Taramon2015} 
that are useful for their intended comparisons with, for example, seismic tomography or plate motions.
The employed numerical resolution is generally low and it is difficult to precisely determine the slab surface
which makes it difficult to use these models for the prediction of the precise thermal structure of the subducting slab
needed to understand the relationships between earthquake locations, slab stratigraphy, and water content.
For this purpose the slab top location should be precisely known and models should have numerical grid spacings of less than a few km in the thermal boundary layers
\cite{vanKeken2002, vankeken2008}.

Finite element models have the particular advantage of being able to precisely prescribe model interfaces (such as the top of the slab
or the Moho of the overriding plate) and to be able to use grid refinement that allows for high resolution near thermal
boundary layers with coarse grids where the velocity and temperature solutions have small gradients, allowing for high 
precision and computational efficiency at the same time
\cite[see, e.g.,][]{PeacockWang1999, WadaWang2009, vanKeken2002, Syracuse2010, vanKeken2019}.
In this paper we will lay the computational groundwork for such high-resolution finite element
models that, due to advances in computational methods and software design, can be used
to study the thermal structure of subduction zones in both 2D and 3D and with time-varying geometry in 
a consistent fashion. This modeling approach will allow us (and other researchers) to study the relationship between
intermediate-depth seismicity, mineralogy, and water content as laid out in, for example, \citet{Wagner2020}.
While the examples presented here are specific for models that
evolve from intermediate dip to flat slab, the open-source modeling framework
we present here is sufficiently general to be used for the thermal modeling
of any deforming slab geometry.

\subsection{Finite element modeling of subduction zones}

Thermal modeling of subduction zones aids in the interpretation of the
chemical and physical processes that take place in the descending slab. To
study the thermal structure of present-day subduction zones with known
geometry and forcing functions (such as age of the oceanic lithosphere at the
trench and convergence speed), 
most existing 2D models \citep[see summary in][]{vanKekenWilson2023} 
combine a kinematically prescribed slab (or slab surface) and a dynamic mantle wedge
(in addition to a dynamic slab if only the slab surface velocity is kinematically prescribed).
This approach works well for regions where the
slab geometry is fixed (even if the forcing functions may change with 
time). The use of finite element methods also allows for the exploration of
3D geometries enabling the study of subduction obliquity,
along-trench variations in slab geometry, and interactions between multiple
slabs \cite{Kneller_vanKeken2008, Bengtson_vanKeken2012, Rosas2016,
WadaHe2017, Plunder2018} that can lead to complicated 3D wedge flow that 
regionally affect the temperature distribution in the subducting slab.

The kinematic-dynamic approach described above has significant limitations
when changes in geometry occur over the lifetime of a subduction
zone. This occurs, for example, when slabs change from intermediate or steep
dip to shallower dip or even to flat slab subduction. 

The thermal structure of flat slabs has been investigated with 2D steady-state
kinematic-dynamic models \cite[e.g.,][]{GutscherPeacock2003, Manea2017} but
the steady-state nature of these models may obscure important effects of 
the geometrical evolution of the slab. An alternative approach is to model
subduction evolution with dynamical models \cite[e.g.][]{Gerya2009,
Liu2022} but the modeled evolution may not conform closely to paleogeographic
constraints and models of slab evolution. It may also be difficult to precisely
trace out the subducting oceanic crust within the evolving slab. The inherent
and complex 3D nature of these regions also suggests the best approach to
understanding the thermal evolution is achieved in a framework that allows for  3D time-dependent
modeling where both geometry and forcing functions can be described. 

Developing such a framework lays forth a number of requirements which extend
beyond the standard \gls{FE} discretization scheme. The subduction zone 
computational model must support: a)~a flexible description of the
time-dependent slab interface geometry; b)~imposition of arbitrary geometry
dependent boundary conditions; and c)~scalable distribution of the discretized
problem for solution with parallel linear algebra packages (i.e., support
efficient computation of small 2D models on local machines and large 3D models
on high performance computers).

To address these requirements we use the components of the FEniCS
project for assembly of our \gls{FE} systems~\cite{fenics:book}. 
The work presented here builds on our extensive experience using FEniCS for
flexible solution of the equations governing subduction zone thermal structure
and mantle dynamics. 
This includes the modeling of the thermal structure of the
subduction slab with and without shear heating \cite{vanKeken2019,abers2020}
and the role of fluid transport through the slab and mantle wedge
\cite{wilson2014,wilson2017,cerpa2017}. We have demonstrated the precision of
the FEniCS applications by comparison to semi-analytical approaches (with
comparisons, for example, to solutions from \citet{MolnarEngland1990} as in
\citet{vanKeken2019}), published subduction zone benchmarks
\cite{vankeken2008}, and intercode comparisons involving detailed
reproductions of published models \cite[e.g.][]{Syracuse2010} that were made
using fully independent finite element software such as Sepran
\cite{vandenBerg2015}. Other useful geodynamical applications using {FEniCS}
include studies of oceanic crust formation and recycling in mantle convection
models \cite{jones2021} and the accurate modeling of buoyancy driven flows in
incompressible and slightly compressible media \cite{sime2020b,sime2022}.

In this paper we will use FEniCS for evolving subduction zone models where 
the final subduction zone geometry is defined by the approximation of a seismically
determined slab surface position (see, e.g., \cref{fig:seismic_data}) by a 
\gls{BSpline}. These \glspl{BSpline} may be manipulated such that a time 
evolving geometry may be defined. The specification of the \gls{BSpline}
further provides convenient interfacing with \gls{CAD} software such that 2D
and 3D volumes of the domain of interest may be generated. These \gls{CAD}
geometries furthermore interface with mesh generators in a straightforward
manner yielding the spatial tessellation of the domain necessary for
discretization of the model by the \gls{FE} method. The nature of the flow on
the slab interface
is
handled by employing Nitsche's method for the weak imposition of boundary
data~\cite{nitsche1971}. This provides us with a much greater flexibility in
the prescribed geometry-dependent flow direction along the slab interface.
Finally, interfacing with the linear algebra solvers provided by the
\gls{PETSc} library gives us the ability to tailor scalable solution methods
for the underlying discretized
\gls{FE} linear system \cite{petsc-web-page}.


In the remainder of this paper we will provide the mathematical and technical
description for this new modeling approach, describe the numerical
implementation, provide examples of the new approach to modeling the evolution
of a flat segment loosely modeled on the Chilean/Argentinian geometry in both 2D
and 3D, and provide a detailed comparison how the 3D models compares to the
simplified 2D cross-sectional model to show that the 3D time-dependent evolution is important
for determining the thermal structure of the subducting crust. 
In a future
article we will use this new modeling ability to specifically test the
hypotheses regarding the cause of intermediate-depth seismicity as laid out in
\citet{Wagner2020}. 


\section{Model}

\subsection{Time evolving domain}

Let $t \in \mathcal{I} = [0, t_\text{slab}]$ be the time domain of the model where
$t_\text{slab}$ is the total time for the slab surface to deform from its
initial to final state. Let $\Omega(t) \subset \mathbb{R}^\mathcal{d}$ be the
spatial domain of interest at a given time $t$, where $\mathcal{d} \in \{ 2, 3
\}$ is the spatial dimension. The domain has boundary $\partial \Omega(t)$
with outward pointing normal unit vector $\hat{\vec{n}}(t)$ and tangential
unit vectors $\hat{\vec{\tau}}_i(t)$, $i=1,\ldots,\mathcal{d}-1$. For brevity
of notation we assume that all quantities deriving from the domain are
functions of time and write $\Omega = \Omega(t)$. Furthermore we uniquely
define each point in $\Omega$ according to the standard Cartesian reference
frame with coordinate tuples and orthogonal unit directions $\vec{x} = (x, z)$
and $(\hat{\vec{x}},
\hat{\vec{z}})$ when $\mathcal{d}=2$ and $\vec{x} = (x, y, z)$ and $(\hat{\vec{x}},
\hat{\vec{y}}, \hat{\vec{z}})$ when $\mathcal{d}=3$. We further define the radial
distance from the origin $r = \Vert \vec{x} \Vert_2$, the unit vector pointing
in the radial direction $\hat{\vec{r}}$, and given the radius of the Earth $r_0$
we define the depth $d = r_0 - r$.

The exterior boundary is divided into components $\partial \Omega_i$ such that
$\partial \Omega = \cup_i \partial\Omega_i$ and no component overlaps $\cap_i
\partial \Omega_i = \emptyset$. On the interior of the geometry we prescribe
an interior boundary, $\Gamma_\text{slab}$, with unit normal
$\hat{\vec{n}}_\text{slab}$. This interior boundary aligns with an
approximation of the subduction zone's slab interface geometry. This interface
defines the surface of bisection of the domain $\Omega$ into
$\Omega_\text{slab}$ and $\Omega_\text{wedge} \cup \Omega_\text{plate}$ such
that $\Omega = \Omega_\text{slab} \cup \Omega_\text{wedge} \cup
\Omega_\text{plate}$ (see \cref{fig:domain}). $\Omega_\text{slab}$ 
extends a depth $d_\text{slab}$ beneath $\Gamma_\text{slab}$
while $\Omega_\text{plate}$ occupies a thickness of $d_\text{plate}$ 
at the top of the domain and above $\Gamma_\text{slab}$.
Each of these subdomains has boundary with outward pointing unit
normal vector $\hat{\vec{n}}_\text{slab}$, $\hat{\vec{n}}_\text{wedge}$ and
$\hat{\vec{n}}_\text{plate}$, respectively.

The interior boundary $\Gamma_\text{slab}$ is further subdivided into
components above a coupling depth, $d_c$, which is embedded in
$\Gamma_\text{slab}$ (a point when $\mathcal{d}=2$ or a curve when
$\mathcal{d}=3$). This coupling depth is the component of $\Gamma_\text{slab}$
below which the slab and wedge velocities will become fully coupled, and above
which a fault discontinuity will be modeled. 
To facilitate this the slab and
wedge domains are separated above $d_c$ such that the slab interface is
labeled $\Gamma_\text{slab fault}$ from the slab side and $\Gamma_\text{wedge
no slip}$ from the wedge side. The specific boundary conditions to be applied
on each of these components will be introduced in \cref{sec:bcs}. A
schematic diagram of an abstract representation of the $\mathcal{d}=2$
subduction zone domain is shown in \cref{fig:domain}. The extrusion of the
shown domain in the $\hat{\vec{y}}$ direction yields a $\mathcal{d}=3$
subduction zone domain. In this case we label the near and far faces
$\partial\Omega_\text{near}$ and $\partial\Omega_\text{far}$, respectively.

\subsection{Underlying \glspl{PDE}}

\begin{table}
\centering
\begin{tabular}{lll}
 Quantity & Symbol & Values, reference values and/or SI units \\ \hline\hline
 Velocity & $\vec{u}$ & \makecell[l]{$u_\text{conv} = \SI{5}{\cm \per \year}$} \\
 Dynamic pressure & $p$ & \si{\pascal} \\
 Temperature & $T$ & \makecell[l]{$T_0 = \SI{273}{\kelvin}$, $T_\text{max} = \SI{1573}{\kelvin}$} \\
 Time & $t$ & $t_\text{slab} = \SI{11}{\mega\year}$ \\
 Position & $\vec{x}$ & \si{\kilo\meter} \\
 Radial distance & $r$ & $r = \Vert \vec{x} \Vert_2$ \\
 Radius of the Earth & $r_0$ & \SI{6371}{\kilo\meter} \\
 Depth & $d$ & $d = r_0 - r$ \\
 Plate depth & $d_\text{plate}$ & $\SI{50}{\kilo\meter}$ \\
 Coupling depth & $d_c$ & $\SI{75}{\kilo\meter}$ \\
 Slab thickness & $d_\text{slab}$ & $\SI{200}{\kilo\meter}$ \\
 Dynamic viscosity & $\eta$ & $\si{\pascal\second}$ (\cref{eq:eta})\\
 Stress tensor & $\sigma$ & \si{\pascal} \\
 Density & $\rho$ &
  $\begin{cases}
   \SI{2700}{\kilo\gram\per\metre\cubed} & \vec{x} \in \overline{\Omega}_\text{plate} \text{ and } \SI{0}{\kilo\meter} \leq d < \SI{40}{\kilo\meter} \\ 
   \SI{3300}{\kilo\gram\per\metre\cubed} & \text{otherwise} 
   \end{cases}$ \\
 Thermal conductivity & $k$ &
        $\begin{cases}
        \SI{2.5}{\watt\per\meter\per\kelvin} & \vec{x} \in \overline{\Omega}_\text{plate} \text{ and } \SI{0}{\kilo\meter} \leq d < \SI{40}{\kilo\meter} \\
        \SI{3}{\watt\per\meter\per\kelvin} & \text{otherwise}
        \end{cases}$ \\
 Heat capacity & $c_p$ & \SI{1250}{\joule\per\kilo\gram\per\kelvin} \\
 Radiogenic heat source & $Q$ &
  $\begin{cases}
   \SI{1.3}{\micro\watt\per\meter\cubed} & \vec{x} \in \overline{\Omega}_\text{plate} \text{ and } \SI{0}{\kilo\meter} \leq d < \SI{15}{\kilo\meter} \\
   \SI{0.27}{\micro\watt\per\meter\cubed} & \vec{x} \in \overline{\Omega}_\text{plate} \text{ and } \SI{15}{\kilo\meter} \leq d \leq \SI{40}{\kilo\meter} \\
   \num{0} & \text{otherwise} \\
   \end{cases}$ \\
 Surface heat flux & $q_\text{surf}$ & \SI{65}{\milli\watt\per\meter\squared}
\end{tabular}
\caption{Summary of physical quantities and reference values used in the model.}
\label{tab:physquants}
\end{table}

The subduction zone evolution is modeled by the incompressible approximation
where we seek velocity $\vec{u} : \Omega \rightarrow \mathbb{R}^\mathcal{d}$, pressure
$p : \Omega \rightarrow \mathbb{R}$ and temperature $T : \Omega \rightarrow
\mathbb{R}$ which satisfy
\begin{linenomath*}
\begin{align}
        -\nabla \cdot \sigma &= \vec{0}, \label{eq:momentum} \\
        \nabla \cdot \vec{u} &= 0, \label{eq:mass} \\
        \rho c_p \del{\dpd{T}{t} + \vec{u} \cdot \nabla T} - \nabla \cdot ( k \nabla T) &= Q, \label{eq:energy}
\end{align}
\end{linenomath*}
with a stress tensor
\begin{linenomath*}
\begin{equation}
\sigma = \eta \del{\nabla \vec{u} + \nabla \vec{u}^\top} - p \vec{I}.
\end{equation}
\end{linenomath*}
Here $\rho$ is the density, $c_p$ is the heat capacity, and $k$ is the thermal
conductivity, which are all considered piecewise constant across the domain. $\vec{I} \in
\mathbb{R}^{\mathcal{d}\times \mathcal{d}}$ is the identity tensor, $\eta : \Omega \rightarrow
\mathbb{R}^+$ is the viscosity, and $Q : \Omega
\rightarrow \mathbb{R}^+ \cup \{ 0 \}$ is the volumetric heat production rate. \Cref{eq:momentum,eq:mass}
comprise the Stokes system of equations conserving momentum and mass,
respectively. \Cref{eq:energy} expresses the conservation of energy of the
system under our incompressible approximation. In our numerical simulations
we solve a rescaled formulation of \cref{eq:momentum,eq:mass,eq:energy}
as described in \cref{sec:rescale}.

The viscosity model employed is that of diffusion creep combined with a near-rigid
crust modeled by high viscosity such that
\begin{linenomath*}
\begin{equation}
\eta(\vec{x}, T) = \begin{cases}
\min \del{\eta_\text{max},  A_\text{diff} \exp \del{ \frac{E_\text{diff}}{R T} }} & \vec{x} \in \Omega_\text{slab} \cup \Omega_\text{wedge}, \\
\num{e5} \eta_\text{max} & \vec{x} \in \Omega_\text{plate}.
\end{cases}
\label{eq:eta}
\end{equation}
\end{linenomath*}
Here $A_\text{diff} = \SI{1.32043e9}{\pascal\second}$ is a constant prefactor,
$E_\text{diff} = \SI{335e3}{\kilo\joule\per\mole}$ is the activation energy,
$R = \SI{8.3145}{\joule\per\mole\per\kelvin}$ is the gas constant and
$\eta_\text{max} = \SI{e26}{\pascal\second}$ is the maximum viscosity within
$\Omega_\text{slab}$ and $\Omega_\text{wedge}$. We are limiting ourselves to this
temperature-dependent rheology that is based on diffusion creep in olivine
\cite{Karato1993}. Detailed comparisons have shown that the near-steady state
thermal structure is very similar to that when olivine dislocation creep
rheology is used \cite{vankeken2008}.



\subsection{Boundary conditions}
\label{sec:bcs}

\begin{table}
\footnotesize
\renewcommand{\arraystretch}{3}
\centering
\begin{tabular}{lll}
 Description & Condition & Location \\ \hline\hline
 Velocity & & \\ \hline
 %
 \makecell[l]{Free slip} & \makecell[l]{$\vec{u} \cdot \hat{\vec{n}} = 0$ \\ and $(\sigma \cdot \hat{\vec{n}}) \cdot \hat{\vec{\tau}}_i = \vec{0}$, \\ $i=1,\ldots,\mathcal{d}-1$} & \makecell[l]{$\partial \Omega_\text{top} \cup \partial \Omega_\text{near} \cup \partial \Omega_\text{far} \cup \partial \Omega_\text{bottom}$}  \\
 Natural in/outlet & $\sigma \cdot \hat{\vec{n}} = \vec{0}$ & \makecell[l]{$\partial \Omega_\text{slab inlet} \cup \partial \Omega_\text{slab outlet} \cup \partial \Omega_\text{wedge outlet}$} \\
 \makecell[l]{Velocity coupled \\ driven slab} & \makecell[l]{$\vec{u} = u_\text{conv} \hat{\vec{\tau}}_\text{conv} + \vec{u}_\text{slab}$} &
 \makecell[l]{$\Gamma_\text{slab} \cup \Gamma_\text{slab fault}$} \\
 %
 %
 Fault zone no slip & $\vec{u} = \vec{u}_\text{slab}$ & $\Gamma_\text{wedge no slip}$ \\
 \hline
 Temperature & & \\ \hline
Surface temperature & $T = T_0$ & $\partial \Omega_\text{top}$ \\
 Inlet slab temperature & $T = T_\text{in}$ & $\partial \Omega_\text{slab inlet}$ \\
 \makecell[l]{Outlet temperature \\ influx} & $T = T_\text{out}$ & \makecell[l]{$\partial \Omega_\text{wedge outlet} \cup \partial \Omega_\text{slab outlet}$ \\ where $\vec{u} \cdot \hat{\vec{n}} < 0$} \\
 \makecell[l]{Outlet temperature \\ outflux} & $k \nabla T \cdot \hat{\vec{n}} = 0$ & \makecell[l]{$\partial \Omega_\text{wedge outlet} \cup \partial \Omega_\text{slab outlet}$ \\ where $\vec{u} \cdot \hat{\vec{n}} \ge 0$} \\
 Zero heat flux & $k \nabla T \cdot \hat{\vec{n}} = 0$ & \makecell[l]{$\partial \Omega_\text{near} \cup \partial \Omega_\text{far} \cup \partial \Omega_\text{bottom}$} \\
 Temperature coupling &
 \makecell[l]{$T_\text{slab} \hat{\vec{n}}_\text{slab} = - T_\text{plate} \hat{\vec{n}}_\text{plate}$ \\
 and $T_\text{slab} \hat{\vec{n}}_\text{slab} = - T_\text{wedge} \hat{\vec{n}}_\text{wedge}$} &
 \makecell[l]{$\Gamma_\text{slab} \cup \Gamma_\text{wedge no slip} \cup \Gamma_\text{slab fault}$}
\end{tabular}
\caption{Boundary conditions imposed in the subduction zone model domain.
See \cref{sec:bcs} regarding specific choices of the functions to impose.}
\label{tab:bcs}
\end{table}

The domain boundary $\partial \Omega$ is subdivided into components as shown
in \cref{fig:domain} (plus $\partial\Omega_\text{near}$ and $\partial\Omega_\text{far}$ 
when $\mathcal{d}=3$). The conditions to be imposed on the velocity and
temperature fields are tabulated in \cref{tab:bcs}. Here we specify the
functions which are imposed.

\subsubsection{Slab convergence and deformation direction}

The down going slab velocity is decomposed into two components, a convergence
speed $u_\text{conv}$ acting in the direction $\hat{\vec{\tau}}_\text{conv}$
and the velocity of the geometry's time evolving deformation,
$\vec{u}_\text{slab}$, such that $\vec{u} = u_\text{conv}
\hat{\vec{\tau}}_\text{conv} + \vec{u}_\text{slab}$ on $\Gamma_\text{slab}
\cup \Gamma_\text{slab fault}$. We define $\hat{\vec{\tau}}_\text{conv}$ in terms
of a prescribed direction, $\hat{\vec{d}}_\text{conv}$. The vector
$\hat{\vec{\tau}}_\text{conv}$ is the unit vector lying tangential to
$\Gamma_\text{slab}$, parallel to $\hat{\vec{d}}_\text{conv}$ and in the
direction of $\hat{\vec{d}}_\text{conv}$. This may also be interpreted as
$\hat{\vec{\tau}}_\text{conv}$ is the vector pointing in direction
$\hat{\vec{d}}_\text{conv}$ and tangential to the intersection of the surface
$\Gamma_\text{slab}$ and the surface defined by the normal vector
$\hat{\vec{r}} \times \hat{\vec{d}}_\text{conv}$ (see \cref{fig:tauconv}). We
therefore define
%
%
\begin{linenomath*}
\begin{align}
\hat{\vec{\tau}}_d &= \hat{\vec{n}}_\text{slab} \times \del{\hat{\vec{r}} \times \hat{\vec{d}}_\text{conv}}, \\
\hat{\vec{\tau}}_\text{conv} &=
\begin{cases}
\hat{\vec{\tau}}_d & \hat{\vec{\tau}}_d \cdot \hat{\vec{d}}_\text{conv} \ge 0, \\
-\hat{\vec{\tau}}_d & \hat{\vec{\tau}}_d \cdot \hat{\vec{d}}_\text{conv} < 0.
\end{cases}
\end{align}
\end{linenomath*}
Furthermore we define the remaining vector lying tangential to $\Gamma_\text{slab}$
and perpendicular to $\hat{\vec{\tau}}_\text{conv}$ and $\hat{\vec{n}}_\text{slab}$.
\begin{linenomath*}
\begin{equation}
        \hat{\vec{\tau}}^\bot_\text{conv} = \hat{\vec{n}}_\text{slab} \times \hat{\vec{\tau}}_\text{conv}.
\end{equation}
\end{linenomath*}
%
%
A schematic diagram of these vectors is shown in \cref{fig:tauconv}.

In addition to the prescribed slab convergence velocity we include the slab
deformation velocity as a result of time-dependent geometry changes. We write
$\vec{u}_\text{slab}$ and $\hat{\vec{d}}_\text{slab}$ to be the slab
deformation velocity and direction, respectively. These quantities will be
fully defined in \cref{sec:disc} where we will also describe the
mathematical representation of $\Gamma_\text{slab}$.

\subsubsection{Temperature models}

The surface temperature is assumed to be a constant value
\begin{linenomath*}
\begin{equation}
T_0 = \SI{273}{\kelvin}.
\end{equation}
\end{linenomath*}
The slab inlet temperature is selected from a half space cooling model
\begin{linenomath*}
\begin{equation}
T_\text{in} = T_0 + (T_\text{max} - T_0) \erf \del{\frac{d}{2
\kappa_\text{in} t_{\SI{50}{\mega\year}}}},
\end{equation}
\end{linenomath*}
where $T_\text{max} = \SI{1573}{\kelvin}$ is the maximum temperature,
$\erf(\cdot)$ is the error function, $\kappa_\text{in} = \left. k /
(\rho c_p) \right|_{\vec{x} \in \partial\Omega_\text{slab inlet}}$ is 
the thermal diffusivity at the slab inlet and $t_{\SI{50}{\mega\year}}$ 
is \SI{50}{\mega\year}.


The outlet temperature is
\begin{linenomath*}
\begin{equation}
	T_\text{out} = \min \left\{ T_\text{1D}, T_\text{max} \right\},
	\label{eq:T_out}
\end{equation}
\end{linenomath*}
where $T_\text{1D}(d)$ is the solution of the initial value problem
\begin{linenomath*}
\begin{equation}
-\frac{\dif}{\dif d} \del{ k \dod{T_\text{1D}}{d}} = Q, \quad T_\text{1D}(d=0) = T_0, \quad \left. \del{k \dod{T_\text{1D}}{d}}\right|_{d=0} = q_\text{surf},
\end{equation}
\end{linenomath*}
where $q_\text{surf}$ is the surface heat flux (see \cref{tab:physquants} for
values used in this work).



\section{Discretization and solution} \label{sec:disc}

In this section we introduce the discretization and solution schemes we employ
to compute numerical approximations of the evolving subduction zone model. We
summarize our procedure in \cref{alg:summary}.

\begin{algorithm}
\begin{algorithmic}
  \State Obtain point clouds from assumed slab geometry (e.g. \cref{fig:seismic_data})
  \State Transform point cloud data to Cartesian coordinates (\cref{sec:transform_data})
  \State Project data to \gls{BSpline}s as final surface approximation (\cref{sec:b_spline_approx})
  \State Generate sequence of domains and meshes of time evolving slab surface (\cref{sec:cad_volume,sec:spacetime_discrete})
  \State Compute initial velocity, pressure and temperature approximation (\cref{sec:picard_solve})
  \ForEach{time step}
    \State Compute velocity, pressure and temperature approximation using \cref{alg:picard} (\cref{sec:picard_solve})
    \State Transfer temperature field to next mesh (\cref{sec:nonmatch_interp})
  \EndFor
\end{algorithmic}
\caption{Computational model summary.}
\label{alg:summary}
\end{algorithm}

\subsection{Mapping slab surface geometries to coordinate data}
\label{sec:transform_data}

Seismic readings provide observations of the slab interface geometry. These
data consist of coordinate tuples of longitude, latitude and depth. Our aim is
to transform these data into a Cartesian system where a central radial
vector aligns with the $\hat{\vec{z}}$ direction.

Let the set of $N_i$ longitude $\lambda_i$, latitude $\mu_i$ and depth
$d_i$ data points for a given point on the slab surface be
\begin{linenomath*}
\begin{equation}
	\Lambda_\text{slab} = \{\lambda_i, \mu_i, d_i \}_{i=1}^{N_i},
\end{equation}
\end{linenomath*}
where longitude and latitude are measured in degrees and depth in kilometers.
Initially these data are transformed to align the central radial vector
with the $\hat{\vec{z}}$ axis of the Cartesian system. We define
\begin{linenomath*}
\begin{equation}
\lambda_\text{mid} = \frac{1}{2}\del{\max_{i=1,\ldots,N_i} \lambda_i +
\min_{i=1,\ldots,N_i} \lambda_i} \text{ and } \mu_\text{mid} =
\frac{1}{2}\del{\max_{i=1,\ldots,N_i} \mu_i + \min_{i=1,\ldots,N_i} \mu_i}
\end{equation}
\end{linenomath*}
such that
\begin{linenomath*}
\begin{equation}
	\overline{\Lambda}_\text{slab} 
	= \{\lambda_i - \lambda_\text{mid}, \mu_i - \mu_\text{mid}, d_i \}_{i=1}^{N_i},
\end{equation}
\end{linenomath*}
with spherical and Cartesian coordinate representations
\begin{linenomath*}
\begin{equation}
	\Phi_\text{slab} = \{r_i, \theta_i, \phi_i\}_{i=1}^{N_i}, \quad
	X_\text{slab} = \{x_i, y_i, z_i\}_{i=1}^{N_i},
\end{equation}
\end{linenomath*}
respectively. The transform to each system is given by
\begin{linenomath*}
\begin{equation}
	\Phi_\text{slab} = \begin{pmatrix}
	r_i \\ \theta_i \\ \phi_i
	\end{pmatrix}
	=
	\begin{pmatrix}
	r_0 - d_i \\
	\mathrm{radians}(\lambda_i - \lambda_\text{mid}) \\
	\mathrm{radians}(90 - (\mu_i - \mu_\text{mid}))
	\end{pmatrix},
	\quad
	X_\text{slab} = \begin{pmatrix}
	x_i \\ y_i \\ z_i
	\end{pmatrix}
	=
	\begin{pmatrix}
	r_i \cos \theta_i \sin \phi_i \\
	r_i \sin \theta_i \sin \phi_i \\
	r_i \cos \phi_i
	\end{pmatrix}.
\end{equation}
\end{linenomath*}

\subsection{Surface data to \gls{BSpline} approximation}
\label{sec:b_spline_approx}

We seek a smooth and continuous approximation of the slab interface surface
from the seismic observation data $X_\text{slab}$. To this end, the data
$X_\text{slab}$ are approximated by an $l_2$ projection to a non-periodic
\gls{BSpline} of order $\mathcal{p}$ \cite[see, e.g.,][]{piegl1997nurbs}.

We define a non-periodic \gls{BSpline} by
\begin{linenomath*}
\begin{equation}
S_{(\mathcal{p},\Xi)}(\xi) = \sum_{i=0}^{n} \vec{C}_i B_{i,\mathcal{p}}(\xi),
\end{equation}
\end{linenomath*}
where $\mathcal{p} = m - n - 1$ is the \gls{BSpline} order, $\vec{C}_i$, $i=0,\ldots,n$, are
the control points, $\Xi = \{ \xi_i \}_{i=0}^{m}$ is the knot vector where each
knot lies in the unit interval $\xi_i \in \sbr{0, 1}$, $i=0,\ldots,m$,
each knot is ordered such that $\xi_i \leq \xi_{i+1}$, $i=0,\ldots,m-1$, and
$B_{i,\mathcal{p}}(\xi)$, $i=0,\ldots,m$, are the \gls{BSpline} basis functions. On the unit
interval $\xi \in \sbr{0, 1}$ these basis functions are
\begin{linenomath*}
\begin{align}
B_{i,0}(\xi) &=
\begin{cases}
1 & \text{if } \xi_i \leq \xi < \xi_{i+1}, \\
0 & \text{otherwise},
\end{cases} \\
B_{i,k}(\xi) &= 
\del{\frac{\xi - \xi_i}{\xi_{i+k} - \xi_i}} B_{i,k-1}(\xi) 
+ \del{\frac{\xi_{i+k+1} - \xi}{\xi_{i+k+1} - \xi_{i+1}}} B_{i+1, k-1}(\xi), \quad k>0.
\end{align}
\end{linenomath*}
A \gls{BSpline} surface of order $\vec{\mathcal{p}} = (\mathcal{p}_1,
\mathcal{p_2})$ is defined by a tensor product of \gls{BSpline}s on the orthogonal
coordinates $\vec{\xi} = (\xi_1, \xi_2) \in \sbr{0, 1}^2$ such that
\begin{linenomath*}
\begin{equation}
S_{(\vec{\mathcal{p}}, \vec{\Xi})}(\vec{\xi}) = \sum_{i=0}^{n_1} \sum_{j=0}^{n_2} \vec{C}_{i,j} B_{i,\mathcal{p}_1}(\xi_1) B_{j,\mathcal{p}_2}(\xi_2).
\end{equation}
\end{linenomath*}

With the definition of the \gls{BSpline} surface in place we define the evolution
of the slab surface with time. Let $\vartheta(t) : \mathcal{I} \rightarrow \sbr{0, 1}$ be
a parametrization of the evolution period of the slab. We write the slab surface
\gls{BSpline}
\begin{linenomath*}
\begin{equation}
S^\text{slab}_{(\vec{\mathcal{p}}, \vec{\Xi})}(\vec{\xi}, t) =
  (1 - \vartheta(t)) S^\text{initial}_{(\vec{\mathcal{p}}, \vec{\Xi})}(\vec{\xi})
+ \vartheta(t) S^\text{final}_{(\vec{\mathcal{p}}, \vec{\Xi})}(\vec{\xi})
\label{eq:slab_evolution}
\end{equation}
\end{linenomath*}
where $S^\text{initial}_{(\vec{\mathcal{p}}, \vec{\Xi})}(\vec{\xi})$ and
$S^\text{final}_{(\vec{\mathcal{p}}, \vec{\Xi})}(\vec{\xi})$ are the initial and final
slab geometries, respectively. We emphasize that $S^\text{initial}$ and
$S^\text{final}$ share a common order, $\vec{\mathcal{p}}$, and knot vector, $\vec{\Xi}$.
Their individual definition is determined by their distinct control points
$\vec{C}^\text{initial}_{i,j}$ and $\vec{C}^\text{final}_{i,j}$.

With appropriate choices of $\vartheta(t)$ the
putative evolution of the slab may be prescribed in the model. In our
experiments we employ a straightforward linear transition such that
\begin{linenomath*}
\begin{equation}
\vartheta(t) =  \frac{t}{t_\text{slab}}.
\end{equation}
\end{linenomath*}
This linear transition also favors a simple
definition of the deformation path undertaken by the modeled slab surface
\begin{linenomath*}
\begin{equation}
\vec{d}_\text{slab} = S^\text{final}_{(\vec{\mathcal{p}}, \vec{\Xi})}(\vec{\xi}) - S^\text{initial}_{(\vec{\mathcal{p}}, \vec{\Xi})}(\vec{\xi})
\end{equation}
\end{linenomath*}
along with the velocity of the slab deformation
\begin{linenomath*}
\begin{equation}
\vec{u}_\text{slab} = \frac{1}{t_\text{slab}} \vec{d}_\text{slab}. 
\end{equation}
\end{linenomath*}

A schematic of the evolution of $S^\text{slab}_{(\vec{\mathcal{p}},
\vec{\Xi})}$ is shown in \cref{fig:slab_evolve}. We highlight that this method
may be extended to arbitrary numbers of prescribed initial, intermediate, and
final subduction zone geometries yielding more sophisticated evolution.

\subsection{Enveloping the slab surface}
\label{sec:cad_volume}

With the representation of the slab evolution using a \gls{BSpline} we now
require its envelopment in a model volume geometry as described in
\cref{fig:domain}. A motivating advantage of a \gls{BSpline} representation of
$\Gamma_\text{slab}$ is its typical compatibility with \gls{CAD} software \cite[e.g.,][]{occ},
which leverage splines to
describe complicated geometries. These splines may be manipulated with a
number of geometric operations, of which we employ:
\begin{itemize}
\item \emph{Extrusion}: Transform a spline along a path generating a higher
dimensional shape from the swept path.
\item \emph{Union, intersection} and \emph{difference}: On a collection
of shapes generate a single shape composed of their boolean union,
intersection or difference, respectively.
\end{itemize}
With these operations we describe the process we employ, in a qualitative
sense, which provides us with the geometry volumes demonstrated in this work.
A diagram of this process is shown in \cref{fig:slab_envelope}.

To generate the slab volume, $\Omega_\text{slab}$, the spline
$S^\text{slab}_{(\vec{p}, \vec{\Xi})}(\vec{\xi}, t)$ is extruded in the
$-\hat{\vec{z}}$ direction by a distance of $d_\text{slab}$. 
To generate the plate and wedge volumes, $\Omega_\text{plate} \cup
\Omega_\text{wedge}$, the spline $S^\text{slab}_{(\vec{p},
\vec{\Xi})}(\vec{\xi}, t)$ is extruded in the $\hat{\vec{z}}$ direction by a
distance greater than the maximum extent of the depth of the spline. This
volume is then intersected with a sphere of radius $r_0$ yielding
$\Omega_\text{plate} \cup \Omega_\text{wedge}$. The distinct volumes
$\Omega_\text{plate}$ and $\Omega_\text{wedge}$ are then formed by embedding
the surface of a sphere of radius $r_0 - d_\text{plate}$. The coupling depth
is also embedded in $S^\text{slab}_{(\vec{\mathcal{p}}, \vec{\Xi})}$ by
finding the intersection with a sphere of radius $r_0 - d_c$.

\subsection{Spatial and temporal discretization}
\label{sec:spacetime_discrete}

The time interval $\mathcal{I} = \sbr{0, t_\text{slab}}$ is discretized into
time steps $\mathcal{I}_{\Delta t} = \cbr{t_0, t_1, \ldots,
t_\text{slab}}$ where $t_0 < t_1 < \ldots < t_\text{slab}$. We write the time
step size $\Delta t_n = t_{n+1} - t_n$ and we use the superscript index $n$ to
denote the evaluation of a function at a particular time step, e.g., $T^n =
T(t_n)$. At each time step, the domain $\Omega^{n} = \Omega(t_n)$ is subdivided into a
tessellation of simplices (triangles when $\mathcal{d}=2$ and tetrahedra when
$\mathcal{d}=3$) which we call a mesh. Each simplex in the mesh is named a
cell and denoted $\kappa$ such that the tessellation $\mathcal{T}^{n} =
\{\kappa^n\}$. The meshing procedure accounts for the internal boundary
$\Gamma_\text{slab}(t_n)$ ensuring facets of cells are aligned
with the surface providing an appropriate approximation.


The spatial components of \cref{eq:momentum,,eq:mass,,eq:energy} are
discretized by the \gls{FE} method. We employ a $P2$-$P1$ Taylor-Hood element pair for
the Stokes system's velocity and pressure approximations 
\cite{taylor1973numerical}
and a standard
quadratic continuous Lagrange element for the temperature
approximation.
The boundary conditions enforced on
$\Gamma_\text{slab fault}$ and $\Gamma_\text{wedge no slip}$ require a
discontinuous velocity solution. Furthermore a discontinuous pressure solution
is required on $\Gamma_\text{slab fault}$, $\Gamma_\text{wedge no slip}$ and
$\Gamma_\text{slab}$. The jump conditions of the temperature approximation are
satisfied by enforcing $C^0$ continuity across $\Gamma_\text{slab fault}$,
$\Gamma_\text{wedge no slip}$ and $\Gamma_\text{slab}$. To this end we define
the following spaces:
\begin{enumerate}
\item $\vec{V}^{h,n} = \{\mathcal{d}$-dimensional piecewise polynomials of degree $2$ defined
in each cell of the mesh $\mathcal{T}^n$ and continuous across cell boundaries
\emph{except} those which overlap the interior boundaries $\Gamma_\text{slab
fault}$ and $\Gamma_\text{wedge no slip}$ $\}$,
\item $Q^{h,n} = \{$scalar piecewise polynomials of degree $1$ defined in each
cell of the mesh $\mathcal{T}^n$ and continuous across cell boundaries
\emph{except} those which overlap the interior boundaries $\Gamma_\text{slab
fault}$, $\Gamma_\text{wedge no slip}$ and $\Gamma_\text{slab}$ $\}$,
\item $S^{h,n} = \{$scalar piecewise polynomials of degree $2$ defined in each
cell of the mesh $\mathcal{T}^n$ and continuous across all cell boundaries$\}$.
\end{enumerate}


On $\mathcal{I}_{\Delta t}$ the time derivative in the the heat equation is
discretized using a finite difference scheme such that
\begin{linenomath*}
\begin{equation}
\dpd{T}{t} \approx \frac{(T^{n+1} - T^{n})}{\Delta t_n}.
\end{equation}
\end{linenomath*}
Using a backward Euler discretization allows us to write the fully discrete \gls{FE} formulation for the model:
find $(\vec{u}^{n+1}_h, p^{n+1}_h, T^{n+1}_h) \in (\vec{V}^{h,n+1} \times
Q^{h,n+1} \times S^{h,n+1})$ such that
\begin{linenomath*}
\begin{align}
	&\mathcal{F}^\text{momentum}(\vec{u}^{n+1}_h, p^{n+1}_h, T^{n+1}_h) = \nonumber \\
	&\quad\quad (\sigma^{n+1}_h, \nabla \vec{v}_h) + A^\text{momentum}_{\partial\Omega}((\vec{u}^{n+1}_h, p^{n+1}_h, T^{n+1}_h), \vec{v}_h) \equiv 0, \label{eq:wf_momentum_fulldiscr} \\
	&\mathcal{F}^\text{mass}(\vec{u}^{n+1}_h, p^{n+1}_h, T^{n+1}_h) = \nonumber \\
	&\quad\quad (\nabla \cdot \vec{u}^{n+1}_h, q_h) + A^\text{mass}_{\partial\Omega}((\vec{u}^{n+1}_h, p^{n+1}_h, T^{n+1}_h), q_h) \equiv 0, \label{eq:wf_mass_fulldiscr} \\
	&\mathcal{F}^\text{energy}(\vec{u}^{n+1}_h, p^{n+1}_h, T^{n+1}_h) = \nonumber \\
	&\quad\quad \del{\rho c_p \del{(T^{n+1}_h - T^n_h) / \Delta t_n + \vec{u}^{n+1}_h \cdot \nabla T^{n+1}_h}, s_h} + (k \nabla T^{n+1}_h, \nabla s_h) \nonumber \\
	&\quad\quad\quad\quad + A^\text{energy}_{\partial\Omega}((\vec{u}^{n+1}_h, p^{n+1}_h, T^{n+1}_h), s_h) - \del{Q, s_h} \equiv 0, \label{eq:wf_energy_fulldiscr}
\end{align}
\end{linenomath*}
for all $(\vec{v}_h, q_h, s_h) \in (\vec{V}^{h,n+1} \times Q^{h,n+1} \times
S^{h,n+1})$. Here $(a, b) = \sum_{\kappa \in \mathcal{T}^{n+1}} \int_\kappa a : b
\dif \vec{x}$ is the inner product on the mesh and the terms
$A^i_{\partial\Omega}(\cdot, \cdot)$ are the terms arising from the weak
imposition of the boundary conditions via Nitsche's method stated in
\cref{sec:bcs}. We refer to \citet{nate2018} regarding the formulation of
these terms. With this discretization scheme we also define the discretized
slab deformation velocity component used in the velocity boundary condition
\begin{linenomath*}
\begin{equation}
\vec{u}^{n+1}_\text{slab}(\vec{x}) = \frac{1}{\Delta t_n} 
\del{S^\text{slab}_{(\vec{p}, \vec{\Xi})}(\vec{\xi}(\vec{x}), t^{n+1}) - S^\text{slab}_{(\vec{p}, \vec{\Xi})}(\vec{\xi}(\vec{x}), t^n)}.
\end{equation}
\end{linenomath*}

\subsection{Nonmatching mesh interpolation}
\label{sec:nonmatch_interp}

An operator $\mathcal{P}(\cdot)$ is necessary to transfer the temperature
field from the previous time step, $T^{n}_h(\mathcal{T}^n)$, to the mesh at
the subsequent time step, $T^n_h(\mathcal{T}^{n+1})$, such that
\begin{linenomath*}
\begin{equation}
T^{n}_h(\mathcal{T}^{n+1}) = \mathcal{P}(T^{n}_h(\mathcal{T}^{n})).
\end{equation}
\end{linenomath*}
The choice of $\mathcal{P}(\cdot)$ must account for cases where subsequent meshes do
not overlap. In this work we design $\mathcal{P}(\cdot)$ to be a
nearest-neighbor interpolation such that $\mathcal{P}(T^{n}_h(\mathcal{T}^{n}))$ interpolates
$T^{n}_h(\mathcal{T}^n)$ in the overlapping volume
$\mathcal{T}^{n+1} \cap \mathcal{T}^n$. In the remaining volume,
$\mathcal{T}^{n+1} \setminus \mathcal{T}^n$, 
$\mathcal{P}(T^{n}_h(\mathcal{T}^{n}))$ interpolates the value of $T^{n}_h(\mathcal{T}^n)$
which lies closest to the interpolation point. Specifically for each
interpolation point $\vec{x}_i$ of $T^{n}_h(\mathcal{T}^{n+1})$ we have
\begin{linenomath*}
\begin{align}
T^n_h(\mathcal{T}^{n+1})(\vec{x}_i) = \mathcal{P}(T^n_h(\mathcal{T}^{n}))(\vec{x}_i) &= 
\begin{cases}
T^n_h(\mathcal{T}^n)(\vec{x}_i) & \forall \vec{x}_i \in \mathcal{T}^n, \\
T^n_h(\mathcal{T}^n)(\argmin_{\vec{y} \in \mathcal{T}^n} (\Vert \vec{x}_i - \vec{y} \Vert)) & \text{otherwise},
\end{cases} \nonumber \\
i &= 1,\ldots,\dim(S^{h,n+1}).
\end{align}
\end{linenomath*}
Our choice of $\mathcal{P}(\cdot)$ here is the motivation for selecting the
backward Euler finite difference scheme in the temporal discretization. A
higher order finite difference scheme will have to carefully account for
fields defined on both $\mathcal{T}^{n+1}$ and $\mathcal{T}^{n}$ in the
\gls{FE} formulation.

\subsection{Picard iteration and computational linear algebra solvers}
\label{sec:picard_solve}

The fully discrete system in
\cref{eq:wf_momentum_fulldiscr,eq:wf_mass_fulldiscr,eq:wf_energy_fulldiscr}
is nonlinear. We use a Picard iterative scheme to compute their solutions' approximations
and minimize the residual formulations. This requires us to split the
solution of the Stokes system from the energy equation. Therefore we introduce
a subscript index, $\ell$, corresponding to the Picard iteration number. Given an
initial guess of the temperature field $T^{n+1}_{h,\ell=0} = \mathcal{P}(T^n_h)$, we compute the sequence
as shown in \cref{alg:picard}.

The linear systems which underlie the \gls{FE} discretization are typically too large
to compute in reasonable time with direct factorization due to the spatial fidelity
required from the mesh. This is especially pertinent in the $\mathcal{d}=3$ case where
computation by direct factorization is unfeasible. We employ an iterative scheme for
both the Stokes and heat equation sub problems in each Picard iteration.
The Stokes system is solved by full Schur complement reduction using
\pgls{FGMRES} iterative method \cite{saad93}. The velocity block is
preconditioned using the algebraic multigrid method with near-nullspace
informed smoothed aggregation as provided by
\pgls{PETSc} \cite{petsc-user-ref}. The pressure block is preconditioned with 
the inverse viscosity weighted pressure mass matrix. The heat equation is
solved using \pgls{GMRES} iterative method \cite{saad86} and preconditioned
with \gls{ILU} factorization. For more details on solving such systems using
iterative schemes and devising appropriate preconditioners see, for example,
\citet{may2008preconditioned}.

\begin{linenomath*}
\begin{algorithm}
\begin{algorithmic}
	\State $\ell=0$
	\State Initialize $T^{n+1}_{h,\ell=0} = \mathcal{P}(T^{n}_h)$
	\Do
		\State $\ell \gets \ell + 1$
		\State Solve for $\vec{u}^{n+1}_{h,\ell}$ and $p^{n+1}_{h,\ell}$: \\
		\hspace{1cm} $\mathcal{F}^\text{momentum}(\vec{u}^{n+1}_{h,\ell}, p^{n+1}_{h,\ell}, T^{n+1}_{h,\ell-1})
		+ \mathcal{F}^\text{mass}(\vec{u}^{n+1}_{h,\ell}, p^{n+1}_{h,\ell}, T^{n+1}_{h,\ell-1}) = 0$
		\State Solve for $T^{n+1}_{h,\ell}$: \\
		\hspace{1cm} $\mathcal{F}^\text{energy}(\vec{u}^{n+1}_{h,\ell}, p^{n+1}_{h,\ell}, T^{n+1}_{h,\ell}) = 0$
	\doWhile{$\Vert T^{n+1}_{h,\ell} - T^{n+1}_{h,\ell-1} \Vert_{L_2} / \Vert T^{n+1}_{h,\ell} \Vert_{L_2} < \mathrm{TOL}$}
\end{algorithmic}
\caption{Picard iterative scheme employed to compute the approximate solution of the
nonlinear system
\cref{eq:wf_momentum_fulldiscr,eq:wf_mass_fulldiscr,eq:wf_energy_fulldiscr}.
Here $\Vert \cdot \Vert_{L_2} = \sqrt{(\cdot, \cdot)}$ is the $L_2$ norm
measure and $\mathrm{TOL} \ll 1$ is the convergence threshold tolerance (selected
to be $\mathrm{TOL} = \num{e-6}$ in this work).}
\label{alg:picard}
\end{algorithm}
\end{linenomath*}

\section{Implementation}

In this section we list the computational tools and libraries which facilitate
our computational model. The \gls{FE} system assembly is enabled by the FEniCS
project, this includes:
\begin{enumerate}
\item Basix for pre-computation of \gls{FE} bases~\cite{scroggs2022},
\item \gls{UFL} for the computational symbolic
	algebra representation of \gls{FE} formulations~\cite{ufl},
\item \gls{FFC} for translation to efficient \gls{FE} kernels~\cite{FFC},
\item DOLFINx for the data structures and algorithms necessary for computing
	\gls{FE} functions, tabulating their degrees of freedom, managing meshes
	and facilitating the solution of \gls{FE} linear systems by third party linear
	algebra packages~\cite{logg:2010}.
\end{enumerate}
The components of the FEniCS project have been demonstrated to be scalable in
the context of thermomechanical analysis in~\citet{richardson2019}, where the
linear operators underlie the momentum and energy \gls{FE} discretizations in
this work also.

DOLFINx-MPC~\cite{dolfinx_mpc} is used in combination with DOLFINx to
construct the function spaces $\vec{V}^{h,n}$, $Q^{h,n}$ and $S^{h,n}$.
Specifically DOLFINx-MPC facilitates strong imposition of equality of the
\gls{FE} functions' degrees of freedom at the $\Gamma_\text{slab}$,
$\Gamma_\text{slab fault}$ and $\Gamma_\text{wedge no slip}$ boundaries
as required by the velocity and temperature boundary conditions.

The Python library NURBS-Python (\texttt{geomdl})~\cite{geomdl} is employed for the \gls{BSpline}
approximation of $\Gamma_\text{slab}$. Its data structures and functions are
necessary for \gls{BSpline} initialization and manipulation along with its
facilitation of the $l_2$ minimization of point cloud positional data to the
\gls{BSpline} surface geometry.

The computational domain is defined using the \gls{CAD} framework offered by
\citet{occ}.
These geometries are then interpreted by
the meshing library \texttt{gmsh}~\cite{gmsh} for generation of the sequence
of simplicial meshes for each time step between the initial and final slab
geometry configurations.

The \pgls{PETSc} library~\cite{petsc-web-page,petsc-user-ref} is used for its
data structures and algorithms facilitating distributed parallel computation
of the linear algebra systems' solutions. This includes the implementations of
the \gls{FGMRES} and \gls{GMRES} methods, along with construction of \gls{ILU}
factorization and construction of algebraic multigrid preconditioners.

Automatic formulation of variational formulation terms arising from the weak
imposition of Dirichlet boundary data in
\cref{eq:wf_momentum_fulldiscr,eq:wf_mass_fulldiscr,eq:wf_energy_fulldiscr} is
provided by \texttt{dolfin\_dg}~\cite{nate2018}. Finally, to build the
necessary environment required to run our model on a high performance computer
we use the Spack package manager \cite{gamblin2015}.

\section{Examples}

Our examples derive their geometric definition of
$S^\text{final}_{(\vec{\mathcal{p}}, \vec{\Xi})}$ from a flat slab geometry
within the subducting Nazca plate shown in \cref{fig:seismic_data}. The
initial slab geometry, $S^\text{initial}_{(\vec{\mathcal{p}}, \vec{\Xi})}$, is
defined by a straight slab dipping at an angle of \SI{30}{\degree} with the
trench aligned with the final state. These initial and final states will
describe the evolution of the slab from the reference frame of a stationary
trench. The $\mathcal{d} = 3$ volumes for each time step are constructed as
described in \cref{sec:cad_volume} with $d_\text{plate} =
\SI{50}{\kilo\meter}$, $d_c = \SI{75}{\kilo\meter}$, and $d_\text{slab} =
\SI{200}{\kilo\meter}$. We choose the slab convergence direction
$\hat{\vec{d}}_\text{conv} = \hat{\vec{x}}$ and speed $u_\text{conv} =
\SI{5}{\centi\meter\per\year}$. The total slab deformation time is
$t_\text{slab} = \SI{11}{\mega\year}$ which is appropriate for the modeled subduction zone \cite{Antonijevic2015}.

From this $\mathcal{d} = 3$ geometry we further form $\mathcal{d} = 2$ slices
by taking cross-sections along the planes defined by constant
$y=\SI{-200}{\kilo\meter}$, $\SI{0}{\kilo\meter}$, and $\SI{200}{\kilo\meter}$.
We seek to compare the $\mathcal{d}=2$ model solutions with corresponding cross
sections of the $\mathcal{d}=3$ results found by post processing.

The splines $S^\text{initial}_{(\vec{p}, \vec{\Xi})}$,
$S^\text{final}_{(\vec{p}, \vec{\Xi})}$ and implicitly
$S^\text{slab}_{(\vec{p}, \vec{\Xi})}$ are defined with order $\vec{p}_i = 2$
and number of control points $n_i + 1 = 8$, $i=1,\ldots,\mathcal{d}-1$. In
each model the meshes are generated with cell size constraints of
\SI{2}{\kilo\meter} within \SI{25}{\kilo\meter} of the velocity coupling depth
$d_c$, \SI{5}{\kilo\meter} along the slab interface $\Gamma_\text{slab}$, and
$\SI{10}{\kilo\meter}$ in the remaining volume. 
This means the nodal point spacing varies from \SI{1} to \SI{5}{\kilo\meter}
in the model.
The degree 2 piecewise
polynomials used for the velocity and temperature function spaces
$\vec{V}^{h,n}$ and $S^{h,n}$ yield a distance between \gls{FE} \gls{DoF}
coordinates of approximately half the cell size. Furthermore in each model the
initial temperature field $T^{n=0}_h$ is prescribed from the computation of
the steady state solution of the nonlinear model, $(\vec{u}^{n=0}_h,
p^{n=0}_h, T^{n=0}_h)$, on the initial mesh $\mathcal{T}^{n=0}$.

\subsection{2D slab}

The temperature and velocity fields computed on the $\mathcal{d}=2$
cross-sections of the $\mathcal{d}=3$ geometry are shown in
\cref{fig:flatslab_temperature_-200,,fig:flatslab_temperature_0,,fig:flatslab_temperature_200},
respectively. Each row corresponds to time snapshots taken over the
$t_\text{slab}=\SI{11}{\mega\year}$ model maximum time which has been
discretized with \num{100} time steps such that $\Delta t =
\SI{0.11}{\mega\year}$. The slab \gls{BSpline} is overlaid in each plot as a dotted
line. Tracers are added in the velocity plots showing pathlines between the
shown time snapshots. Should a tracer leave the geometry between each
snapshot, it is removed from the visualization leaving only its remaining
tail. The geometry deformation is not shown
between snapshots and the tracers do not cross over $\Gamma_\text{slab}$ at
any time in the simulation (though their pathlines may appear to do so).


Convergence of the surface temperature as a function of time step
size in the temporal discretization is shown in
\cref{fig:flatslab_temperature_convergence}. In
\cref{fig:flatslab_temperature_steady_vs_time} we show the slab surface
temperature as a function of depth as computed in the steady state on the
initial geometry $S^\mathrm{initial}_{(\vec{\mathcal{p}},
\vec{\Xi})}$, as computed in the full time dependent $\mathcal{d}=2$ model at
time $t = t_\text{slab}$ and as the steady state on the final geometry
$S^\mathrm{final}_{(\vec{\mathcal{p}}, \vec{\Xi})}$. Finally, the temperature
as a function of depth along $\Gamma_\text{slab}$ at the final time
$t=t_\text{slab}$ are shown in \cref{fig:flatslab_temperature_2d3d}.

\subsection{3D slab}

Snapshots of the $\mathcal{d} = 3$ case temperature and velocity
approximations evaluated on the slab interface are shown in
\cref{fig:2a_surfaces_pushforward}. As in the $\mathcal{d}=2$ case we
discretize the temporal domain with \num{100} time steps such that $\Delta t =
\SI{0.11}{\mega\year}$. Overlaid on the velocity plot are arrows indicating
the direction of flow on the surface in the $\hat{\vec{x}}$ and
$\hat{\vec{y}}$ directions along with cross markers with size corresponding to
normalized speed in the $-\hat{\vec{z}}$ direction (into the page). 
Cross-sections of the temperature and velocity solution at 
$y = \SI{-200}{\kilo\meter}$, $\SI{0}{\kilo\meter}$, and $\SI{200}{\kilo\meter}$ are
shown in
\cref{fig:3d_slice_flatslab_temperature_-200,,fig:3d_slice_flatslab_temperature_0,,fig:3d_slice_flatslab_temperature_200},
respectively. Tracers and their pathlines are not added to these velocity
cross-sections due to the inability to visualize their $\hat{\vec{y}}$
component.

Cross-sections of the temperature field taken at constant 
$y = \SI{-200}{\kilo\meter}$, $\SI{0}{\kilo\meter}$, and $\SI{200}{\kilo\meter}$ as
a function of depth along $\Gamma_\text{slab}$ at final time $t=t_\text{slab}$
is shown in \cref{fig:flatslab_temperature_2d3d}. These data overlay the slab
surface temperatures as a function of depth computed from the corresponding
$\mathcal{d} = 2$ models. Furthermore, convergence of the surface temperatures
as a function of time step size in the temporal discretization at these cross
sections is shown in \cref{fig:flatslab_temperature_convergence}. The volume
of the $\mathcal{d}=3$ model at $t=t_\text{slab}$ with these cross-sections
and additional path tracers is shown in \cref{fig:flatslab_3draytrace}.

\subsection{Discussion}

The surface temperatures computed from the $\mathcal{d} = 2$ and
$\mathcal{d}=3$ models shown in \cref{fig:flatslab_temperature_2d3d} indicate
a warming of the slab above the coupling point. This appears to be caused by
the slab surface transitioning to a shallower angle than the initial
condition, pushing the surface into a warmer region of the wedge (see
\cref{fig:flatslab_temperature_-200,fig:flatslab_temperature_0,fig:flatslab_temperature_200,fig:3d_slice_flatslab_temperature_-200,fig:3d_slice_flatslab_temperature_0,fig:3d_slice_flatslab_temperature_200}).
Examining the $y = \SI{200}{\kilo\meter}$ cross-section in
\cref{fig:3d_slice_flatslab_temperature_200} the slab surface does not evolve
to a much shallower depth as in the $y = \SI{-200}{\kilo\meter}$ and $y =
\SI{0}{\kilo\meter}$ cross-section cases. This corresponds with the less
significant warming of the slab surface above $d_c$. Consider also the slab
surface temperatures shown in \cref{fig:flatslab_temperature_steady_vs_time}
which increase as the slab evolves from the initial steady state to $t =
t_\text{slab}$, and cools when evolved to the steady state with no further
slab deformation.

In all cases the steady state solution used for the initial temperature field
$T_h^{n=0}$ exhibits a diffusive thickening of the plate within the
approximate depths of $\SI{50}{\kilo\meter}$ and $\SI{100}{\kilo\meter}$. This
feature persists through the simulation and is displaced by the slab deformation.
Future models may be improved by prescribing the initial temperature field
computed from an unsteady simulation run to a time just after transient effects
become negligible.

The configuration of the slab surface temperature in the $\mathcal{d} = 3$
case shown in \cref{fig:2a_surfaces_pushforward} is largely dictated by the
velocity boundary condition applied to $\Gamma_\text{slab}$. Choosing
$\hat{\vec{d}}_\text{conv} = \hat{\vec{x}}$ restricts the velocity profile to
be very similar to the $\mathcal{d} = 2$ cases along $\Gamma_\text{slab}$.
Deviating from this decision, one avenue is to choose
$\hat{\vec{d}}_\text{conv} = -\hat{\vec{z}}$ which would yield a convergence
velocity in the direction of steepest descent. However in this case, the flow
above and below $\Gamma_\text{slab}$ will become unreasonable for the
subduction zone model as a result of satisfying mass conversation, $\nabla \cdot
\vec{u} = 0$. These flows, which are not realistic in a subduction zone model,
typically form as velocity fields impinging or jetting out from
$\Gamma_\text{slab}$ in order to account for diverging and converging flows on
the $\Gamma_\text{slab}$ topology, respectively.
An approach to alleviate this issue is to solve for some component of the
flow, $\vec{u}$, on $\Gamma_\text{slab}$ implicitly. For example, the velocity
prescription on $\Gamma_\text{slab}$ could be changed such that only the
$\hat{\vec{\tau}}_\text{conv}$ component is imposed allowing the remaining
tangential component in the $\hat{\vec{\tau}}^\bot_\text{conv}$ direction to
be implicit in the model. This however introduces an issue where the
deformation velocity component, $(\vec{u}_\text{slab} \cdot
\hat{\vec{\tau}}^\bot_\text{conv}) \hat{\vec{\tau}}^\bot_\text{conv}$, must be
neglected. 
Another approach would be to impose a convergence velocity
$\vec{u}_\text{conv}$, such that $\vec{u}|_{\Gamma_\text{slab}} =
\vec{u}_\text{conv} + \vec{u}_\text{slab}$, which is computed from a Stokes
problem of topological dimension $\mathcal{d} - 1$ defined on
$\Gamma_\text{slab}$. The divergence free constraint defined on the topology
of the surface would then ensure no regions of converging or diverging flow.
However, the complexity of the mathematical formulation of this problem as well as its
implementation for parallel computation is challenging.

The slab temperature approximation close to $\partial\Omega_\text{slab
outlet}$ and $\partial\Omega_\text{wedge outlet}$ is significantly affected by
the non-overlapping component of the interpolation operation
$\mathcal{P}(\cdot)$ described in \cref{sec:nonmatch_interp}. This is
indicated, for example, in the $t>0$ cases of
\cref{fig:flatslab_temperature_2d3d} at depths below \SI{215}{\kilo\meter}
(i.e. the component of the $\Gamma_\text{slab}$ closest to
$\partial\Omega_\text{slab outlet}$ and $\partial\Omega_\text{wedge outlet}$).
One can see a small downturn in the temperature which arises from
interpolation of $T_\text{out}$ (\cref{eq:T_out}) which is colder than the
material in the volume which is displaced by the moving slab. This issue could
be addressed by ensuring all meshes overlap such that the overall volume
remains consistent negating the need for non-overlapping interpolation.
However, this would introduce a large computational cost resolving a volume
which is largely spatially removed from the domain of interest close to
$\Gamma_\text{slab}$.

The decision to choose the \gls{BSpline} properties $\vec{p}_i = 2$ and $n_i + 1 =
8$, $i=1,\ldots,\mathcal{d}-1$, was made to balance production of a robust
numerical model against the performance of iterative solvers applied to the
linear system underlying the Stokes problem. Choosing a greater fidelity in
the knot vector lead to degradation of the rate of convergence of the
\gls{FGMRES} method for seismic data which exhibit rapid non-smooth changes in
the slab geometry. Future development of the model would investigate methods
to retain the robust solution of the velocity and pressure approximations with
greater spatial fidelity of $\Gamma_\text{slab}$. Additionally the geometric
operations required to define the volume using \gls{CAD} as described in
\cref{sec:cad_volume} become prohibitively expensive as the \gls{BSpline}
approximation order and control point vectors' cardinality increases.

We finish discussion on the caution that these models are in sensu stricto based
on a toy model (if admittedly a complicated one). The results presented here should
be interpreted to indicate that precise description of the slab evolving geometry
leads to significant differences between 3D models and 2D cross-sections, but
the temperature-pressure paths should not be used to compare directly to existing
slab models or observations of flat slab subduction. In future work we will apply
this modeling frame work to regions of flat slab subduction with locally adjusted
parameters for geometry, coupling point, structure of the overriding plate, etc.


\section{Conclusion}

We have devised, implemented, and demonstrated a numerical model of a
subduction zone which accounts for a kinematic prescription of a geometry
evolving slab surface. We do this by approximating seismic observation of slab
geometries with a \gls{BSpline}. By constructing a deformation path for the
\gls{BSpline} surface from an initial to a final slab geometry, we are able to
evolve this prescribed slab surface geometry over time. Enveloping the slab
surface spline in a volume using \gls{CAD} allows us to create a sequence of
meshes in which we compute approximations of velocity, pressure and
temperature of a subduction zone model discretized by the \gls{FE} method.

\appendix
\section{Model equations rescaling}
\label{sec:rescale}

Using velocity, length and viscosity scales $u_r$,
$h_r$ and $\eta_r$, respectively we define the rescaled
quantities
\begin{align}
      \vec{u} &= u_r \vec{u}^\prime,
       \quad \vec{x} = h_r \vec{x}^\prime,
       \quad \eta = \eta_r \eta^\prime, \nonumber \\
       \nabla &= \frac{1}{h_r} \nabla^\prime,
       \quad t = \frac{h_r}{u_r} t^\prime,
       \quad k = h_r u_r k^\prime,
       \quad Q = \frac{u_r}{h_r} Q^\prime.
\end{align}
Employing these quantities we arrive at the rescaled Stokes-energy formulation
of \cref{eq:momentum,eq:mass,eq:energy} where we seek $\vec{u}^\prime$, $p^\prime$
and $T^\prime$ such that
\begin{align}
      -\eta_r \frac{u_r}{h_r^2}\nabla^\prime \cdot \sigma^\prime &= \vec{0}, \\
      \frac{u_r}{h_r} \nabla^\prime \cdot \vec{u}^\prime &= 0, \\
      \rho c_p \left( \frac{u_r}{h_r} \dpd{T}{{t^\prime}} + \frac{u_r}{h_r} \vec{u}^\prime \cdot \nabla^\prime T \right)
       - \frac{h_r u_r}{h_r^2} \nabla^\prime \cdot \left( k^\prime \nabla^\prime T \right) &= \frac{u_r}{h_r} Q^\prime,
\end{align}
which after simplification reads
\begin{align}
      - \nabla^\prime \cdot \sigma^\prime &= \vec{0}, \\
      \nabla^\prime \cdot \vec{u}^\prime &= 0, \\
      \rho c_p \left(\dpd{T}{{t^\prime}} + \vec{u}^\prime \cdot \nabla^\prime T \right)
       - \nabla^\prime \cdot \left( k^\prime \nabla^\prime T \right) &= Q^\prime.
\end{align}
The numerical values used in our computations are $h_r = \SI{1}{\kilo\meter}$,
$u_r = \SI{5}{\centi\meter\per\year}$ and $\eta_r = \SI{e21}{\pascal}$.


\begin{backmatter}

\section*{Abbreviations}

B-spline: B\'ezier spline; CAD: Computer-Aided Design; DoF: Degree of Freedom; FE: Finite element;
FFC: FEniCS Form Compiler; FGMRES: Flexible Generalized Minimal Residual; GMRES: Generalized Minimal Residual;
iLU: incomplete LU; PDE: Partial Differential Equation; PETSc: Portable Extensible Toolkit for Scientific computation;
UFL: Unified Form Language.

\section*{Availability of data and material}

An implementation of the subduction zone model is provided at
\url{https://github.com/nate-sime/mantle-convection}.

The data generated by the subduction zone model code and presented in this
paper is available in \cite{sime2023data}.

\section*{Competing interests}

The authors declare that they have no competing interest.

\section*{Funding}

This work was supported by National Science Foundation grant 2021027.

\section*{Authors' contributions}

NS developed and tested the FEniCS finite element approach in collaboration with CW. PvK aided in
benchmarking the 2D subduction zone models. All contributed to writing the manuscript.

\section*{Acknowledgements}

The authors thank Lara Wagner for providing the slab surface
geometry used in \cref{fig:seismic_data} and for discussions. The authors also thank
J{\o}rgen Dokken for his advice regarding the use of DOLFINx-MPC.

\section*{Endnotes}


%

\bibliographystyle{peps-art} 
\bibliography{references_PEPS}  


\newpage
\begin{figure}
\centering
\begin{subfigure}{.75\textwidth}
  \centering
  \includegraphics[width=0.9\textwidth]{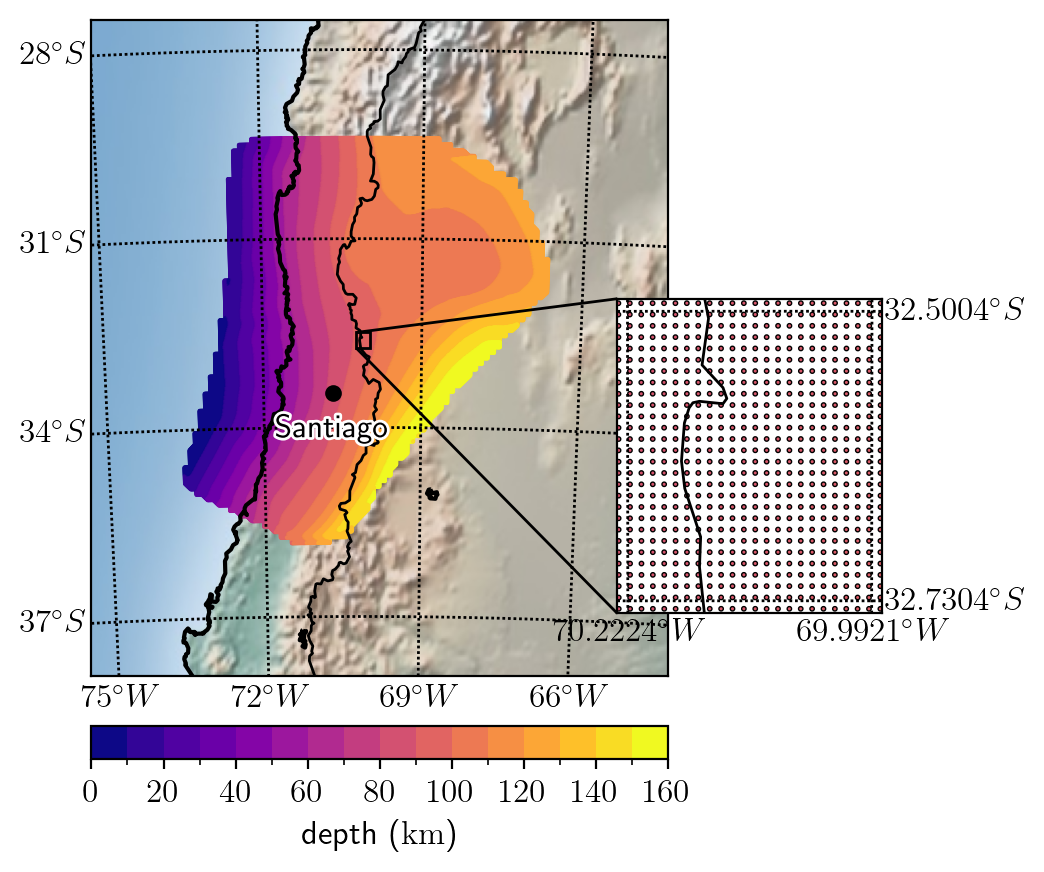} \\
\end{subfigure}
%
\caption{
    Subduction zone interface geometry data points as retrieved from
    seismological observations. The resolution of the data is highlighted along
    the border of Chile and Argentina. Geometry of the present-day structure
    of the flat slab segment is based on \protect\citet{Anderson2007}.}
\label{fig:seismic_data}
\end{figure}

\begin{figure}
\centering
\includegraphics[width=1.\linewidth]{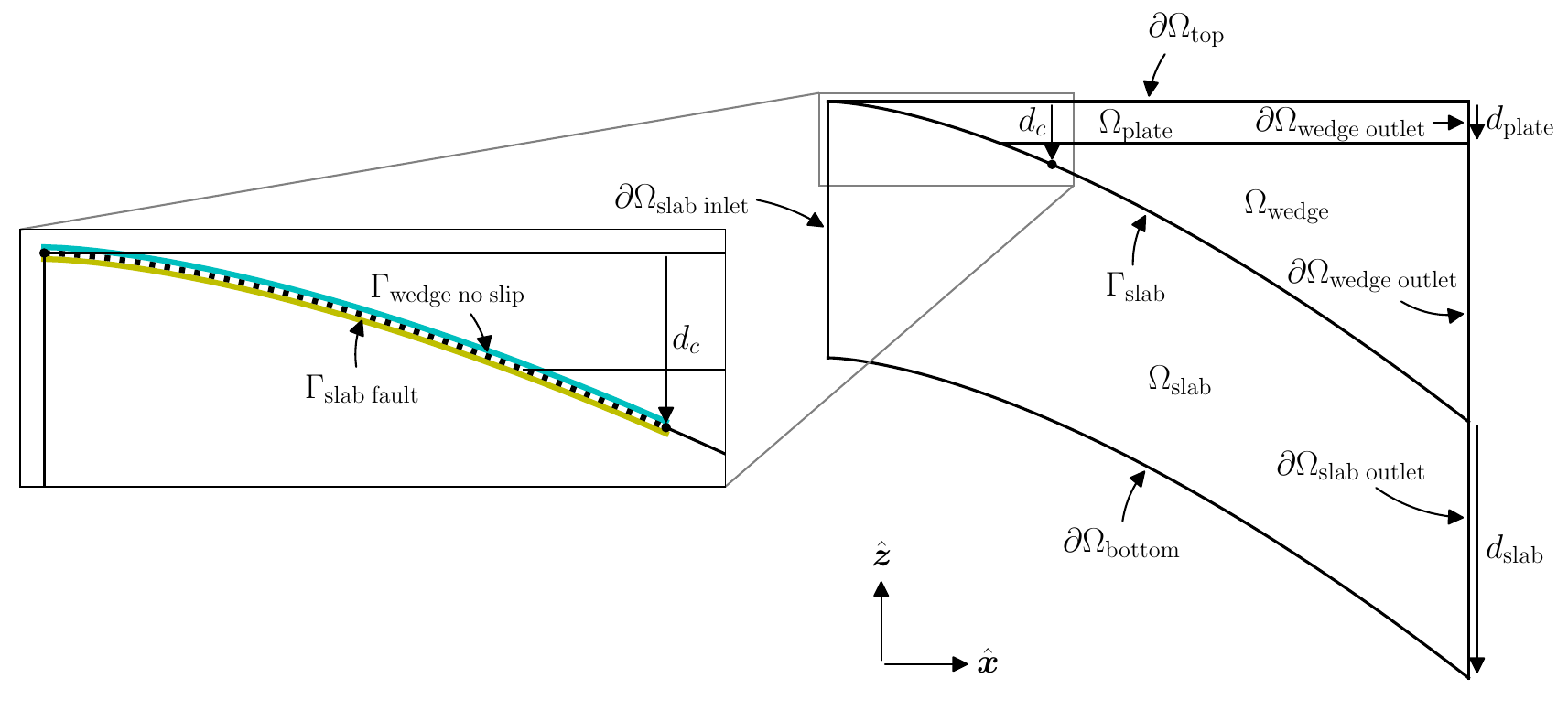}
\caption{
    Schematic diagram of the subduction zone geometry employed in the model when $\mathcal{d}=2$. The
    inset axis highlights the subdivision of the slab interface above the coupling depth
    $d_c$. Extension to $\mathcal{d}=3$ can be envisaged as the extrusion of this
    geometry in the $\hat{\vec{y}}$ direction.}
\label{fig:domain}
\end{figure}

\begin{figure}
\centering
\includegraphics[width=.5\linewidth]{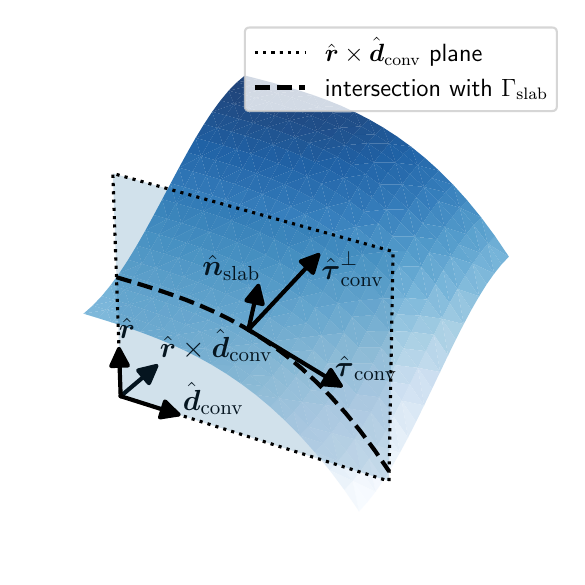}
\caption{Schematic of the slab convergence direction. The surface
$\Gamma_\text{slab}$ is colored by depth for visual aid. The plane defined by
$\hat{\vec{r}} \times \hat{\vec{d}}_\text{conv}$ and its intersection with
$\Gamma_\text{slab}$ are highlighted. The slab surface normal vector
$\hat{\vec{n}}_\text{slab}$, the slab convergence direction $\hat{\vec{\tau}}_\text{conv}$ and the
perpendicular vector $\hat{\vec{\tau}}^\bot_\text{conv}$ are drawn as arrows on the
surface.}
\label{fig:tauconv}
\end{figure}

\begin{figure}
\centering
\includegraphics[width=1.\linewidth]{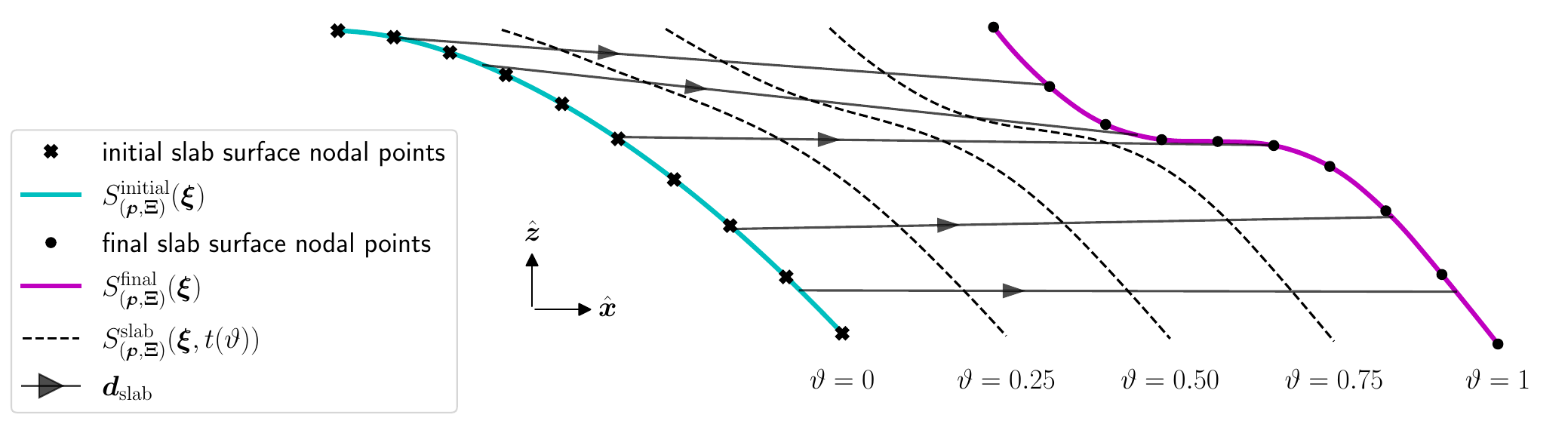}
\caption{
    Schematic diagram of the slab interface geometry designed as a \gls{BSpline} in
    physical space, $(x, z)$, evolving over parametrized time, $\vartheta$.
    The putative initial and observed final coordinates of points on the assumed slab surface are interpolated to
    splines with an equal number of control points and equal number of knot vectors
    yielding $\Gamma_\text{slab}(\vartheta=0)$ and
    $\Gamma_\text{slab}(\vartheta=1)$. The slab interface
    surface at intermediate time is generated according to \cref{eq:slab_evolution}. The
    displacement vector of the slab surface, $\vec{d}_\text{slab}$, is shown
    from the initial to the final configuration.}
\label{fig:slab_evolve}
\end{figure}

\begin{figure}
\centering
\includegraphics[width=.45\linewidth]{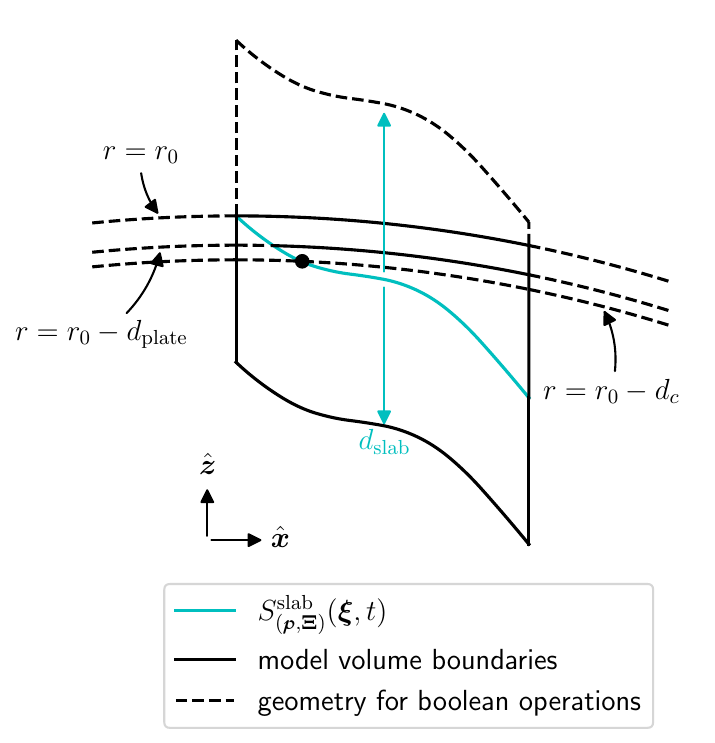}
\caption{
    Schematic of the slab surface envelopment into a volume by employing
    spline manipulations typically offered by \gls{CAD} software.}
\label{fig:slab_envelope}
\end{figure}

\begin{figure}
\centering
\includegraphics[width=.85\linewidth]{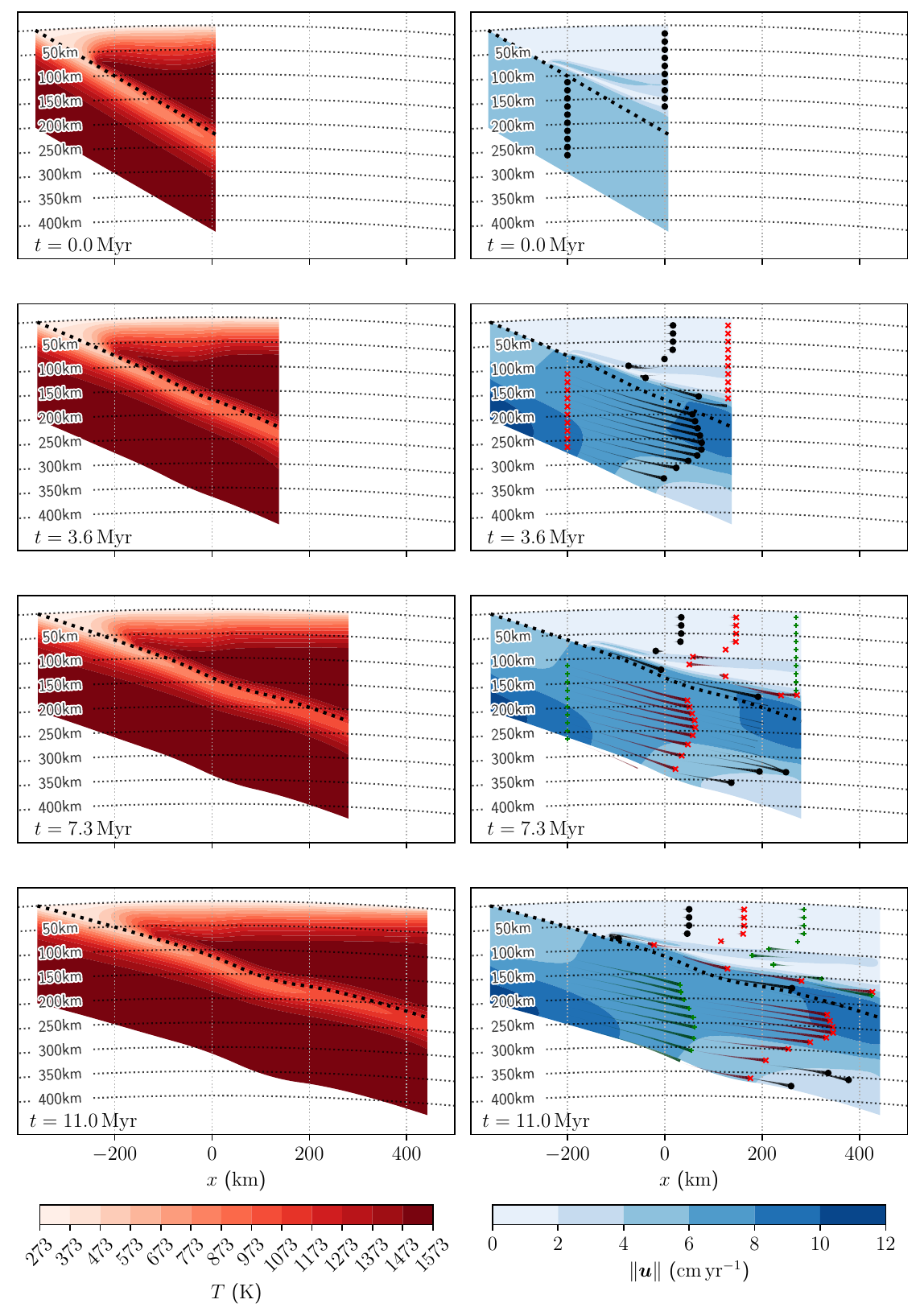}
\caption{Temperatures and speeds around the evolving slab over \SI{11}{\mega\year}.
The model is solved with $\mathcal{d}=2$ where the geometry is taken from a
cross-section of the $\mathcal{d} = 3$ volume parallel to the $y$-axis at location
$y=\SI{-200}{\kilo\meter}$. The time interval is discretized with \num{100}
time steps such that $\Delta t = \SI{0.11}{\mega\year}$. Tracers are added in 
the speed plots showing pathlines between the time snapshots.}
\label{fig:flatslab_temperature_-200}
\end{figure}

\begin{figure}
\centering
\includegraphics[width=.85\linewidth]{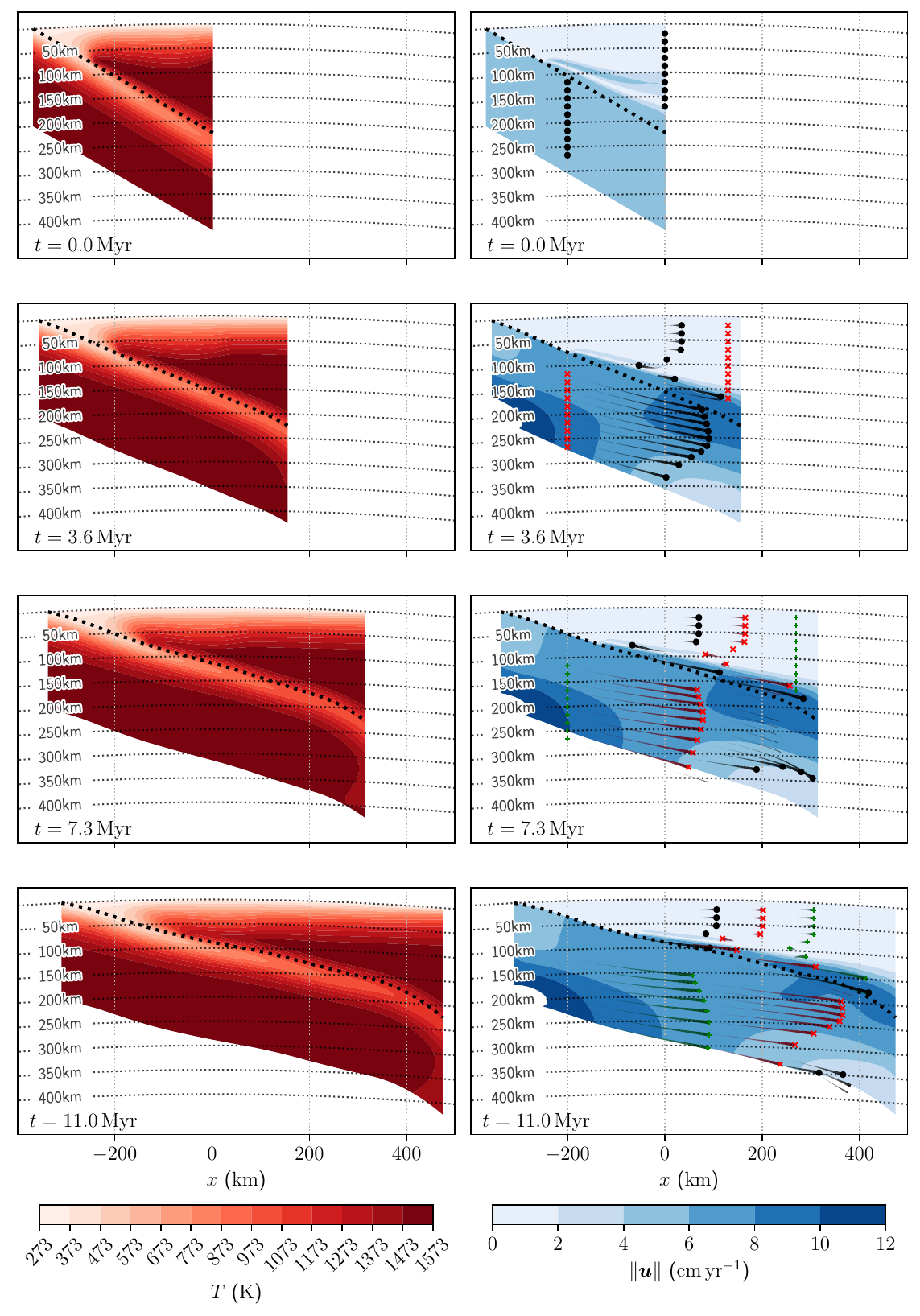}
\caption{As Figure \ref{fig:flatslab_temperature_-200} but now for $y=\SI{0}{\kilo\meter}$.}
\label{fig:flatslab_temperature_0}
\end{figure}

\begin{figure}
\centering
\includegraphics[width=.85\linewidth]{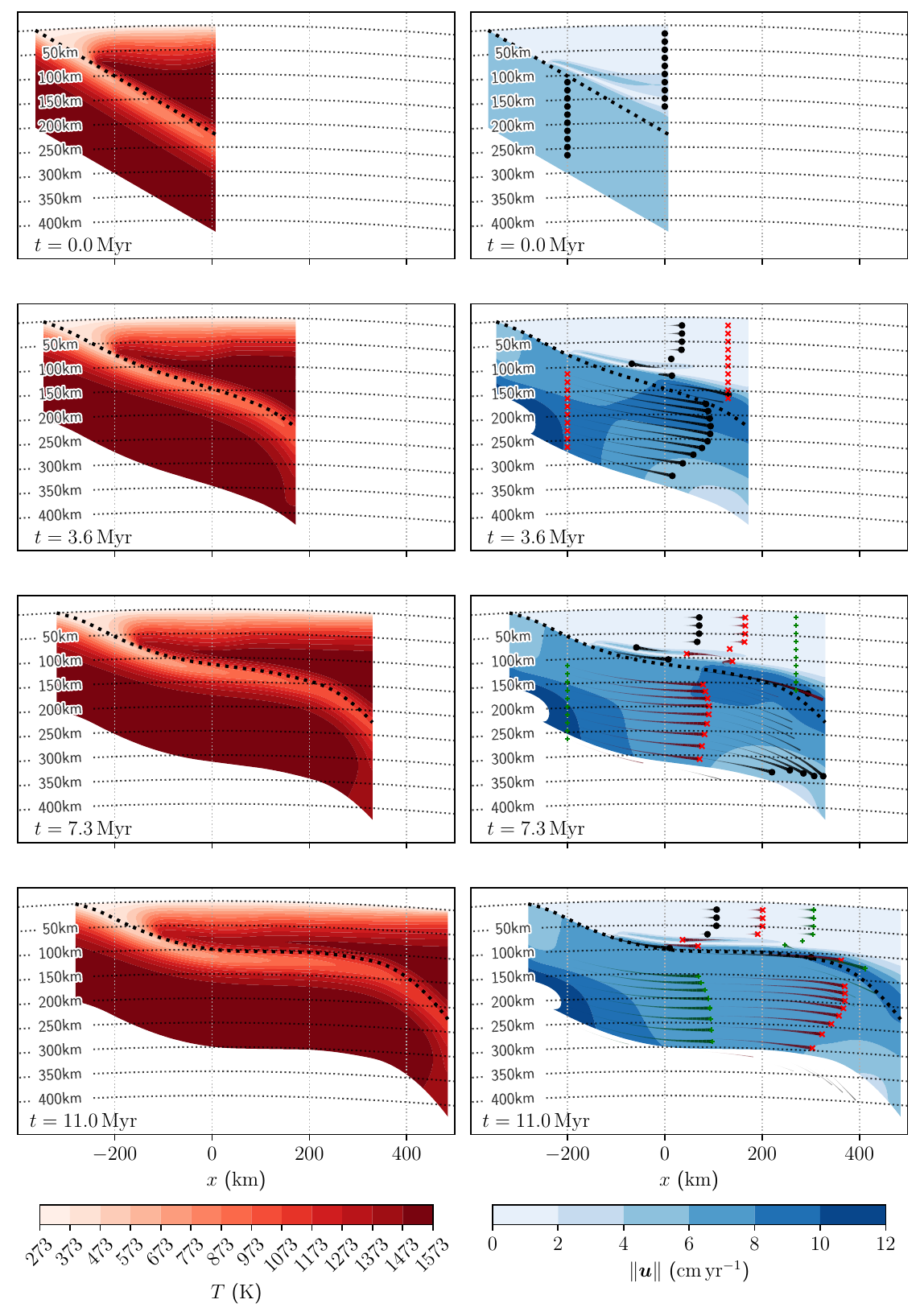}
\caption{As Figure \ref{fig:flatslab_temperature_-200} but now for $y=\SI{200}{\kilo\meter}$.}
\label{fig:flatslab_temperature_200}
\end{figure}

\begin{figure}
\centering
\includegraphics[width=1.\linewidth]{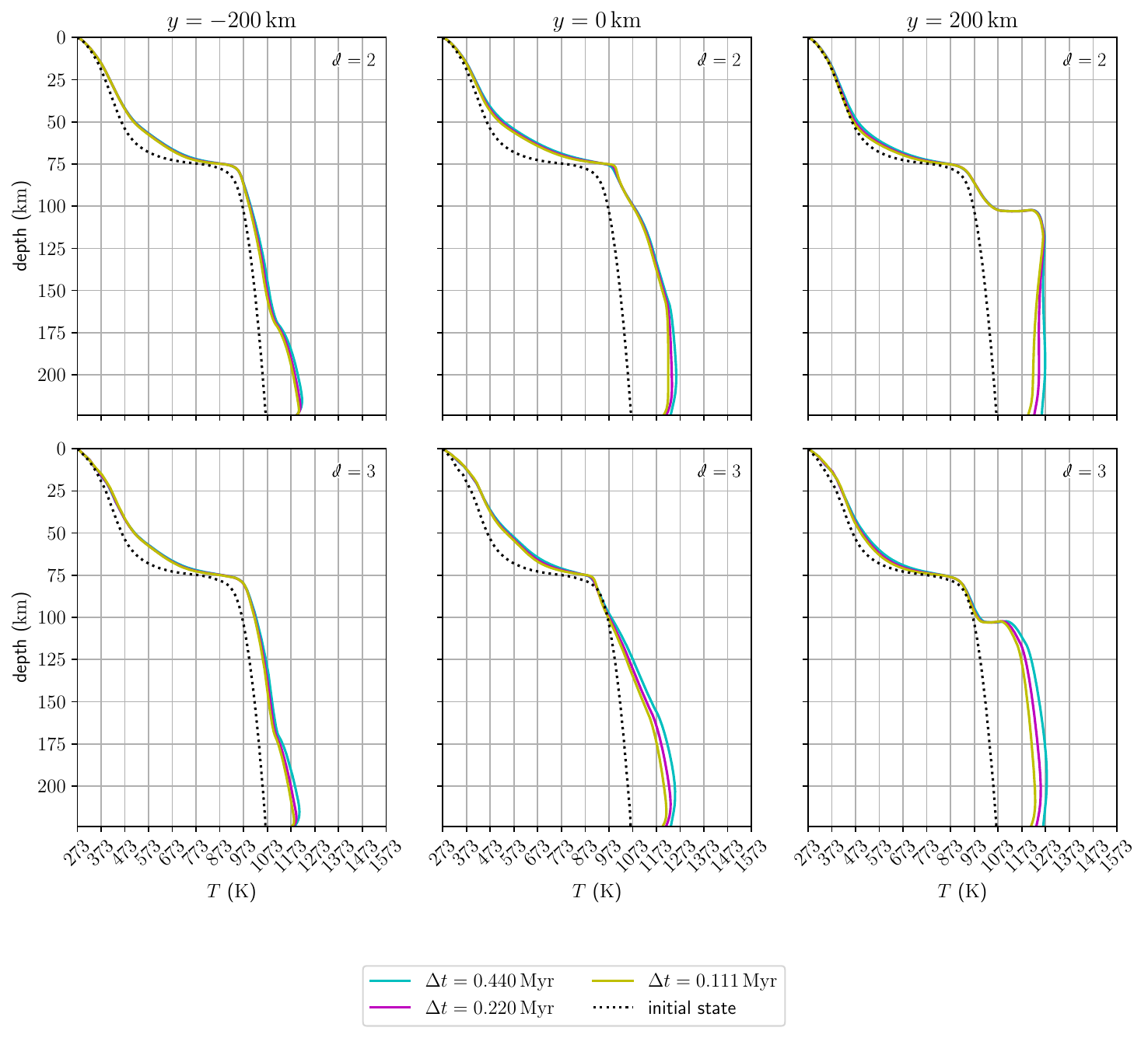}
\caption{Temporal convergence test of the $\mathcal{d}=2$ models shown in
\cref{fig:flatslab_temperature_-200,fig:flatslab_temperature_0,fig:flatslab_temperature_200}
and $\mathcal{d}=3$ models shown in \cref{fig:3d_slice_flatslab_temperature_-200,fig:3d_slice_flatslab_temperature_0,fig:3d_slice_flatslab_temperature_200}.
Here, in each case, we plot the surface temperatures as a function of depth
along $\Gamma_\text{slab}$ at final time $t(\vartheta=1) =
\SI{11}{\mega\year}$. We also overlay the initial temperature along the slab
at $t = 0$ as a dotted line. For each case the temporal domain is discretized by
\num{25}, \num{50} and \num{100} time steps yielding the time step
sizes shown in the legend.}
\label{fig:flatslab_temperature_convergence}
\end{figure}

\begin{figure}
\centering
\includegraphics[width=1.\linewidth]{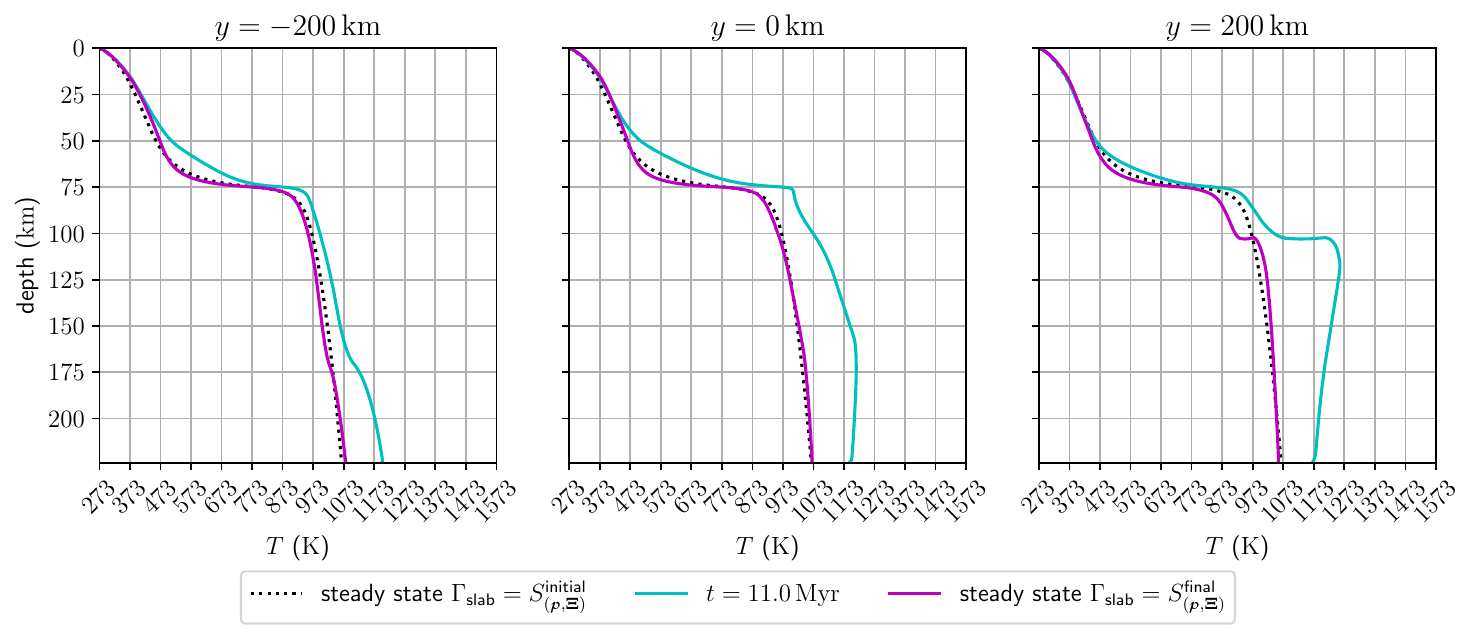}
\caption{Comparison of the $\mathcal{d} = 2$ model slab surface
temperatures as a function of depth in the steady state on the initial
geometry, time dependent $t=\SI{11}{\mega\year}$ and steady state on the final
geometry.}
\label{fig:flatslab_temperature_steady_vs_time}
\end{figure}

\begin{figure}
\centering
\includegraphics[width=0.8\linewidth]{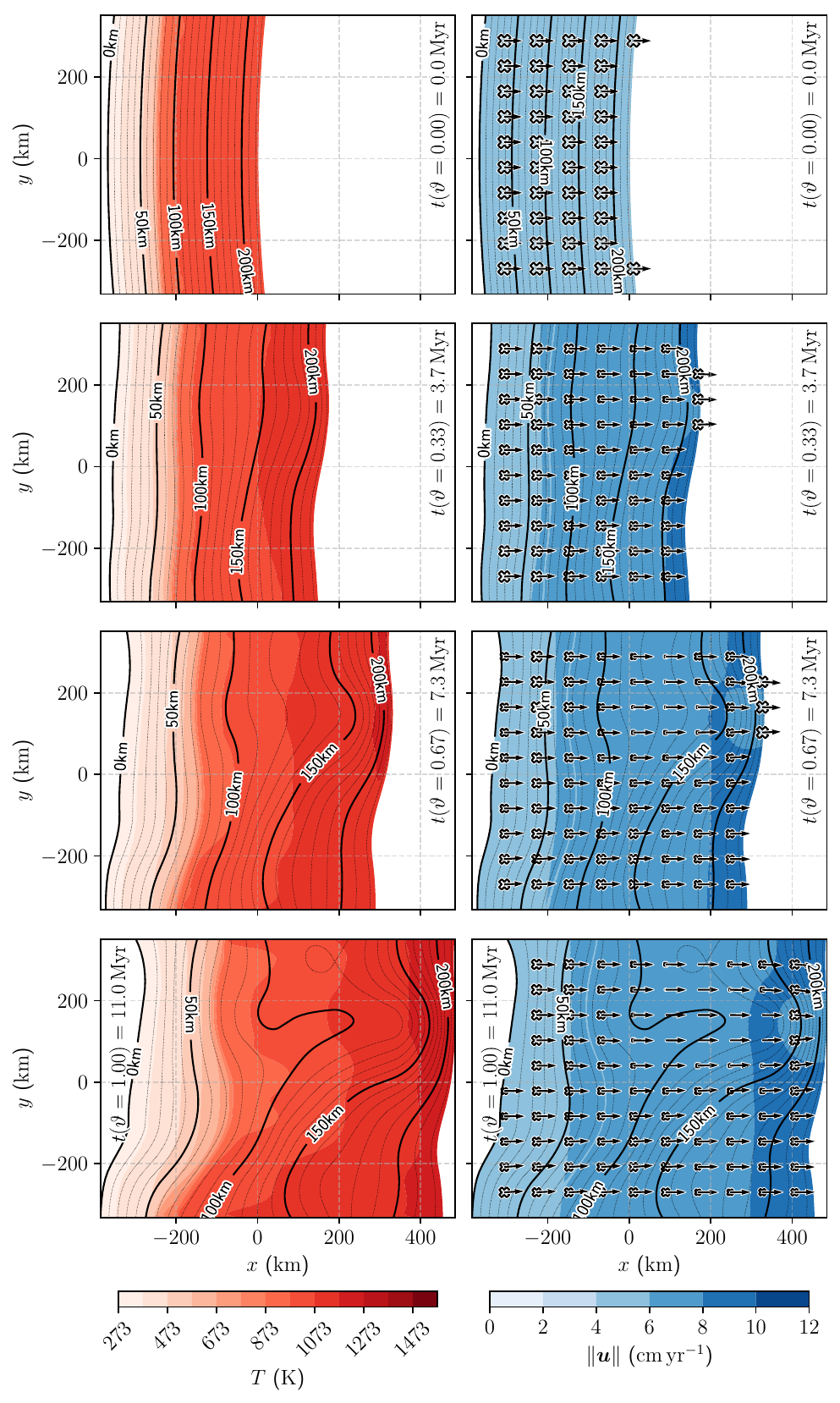}
\caption{Computed $\mathcal{d}=3$ model slab surface temperatures and
velocities with an initial \SI{30}{\degree} straight dip evolving into the
Nazca slab geometry over \SI{11}{\mega\year} (cf.~\cref{fig:seismic_data}).
The time interval is discretized with \num{100} time steps such that $\Delta t
= \SI{0.11}{\mega\year}$. In the instantaneous velocity plot, crosses
represent velocity in the $-\hat{\vec{z}}$ direction (into the page) and
arrows correspond to the velocity component acting in the $\hat{\vec{x}}$ and
$\hat{\vec{y}}$ directions.}
\label{fig:2a_surfaces_pushforward}
\end{figure}

\begin{figure}
\centering
\includegraphics[width=.85\linewidth]{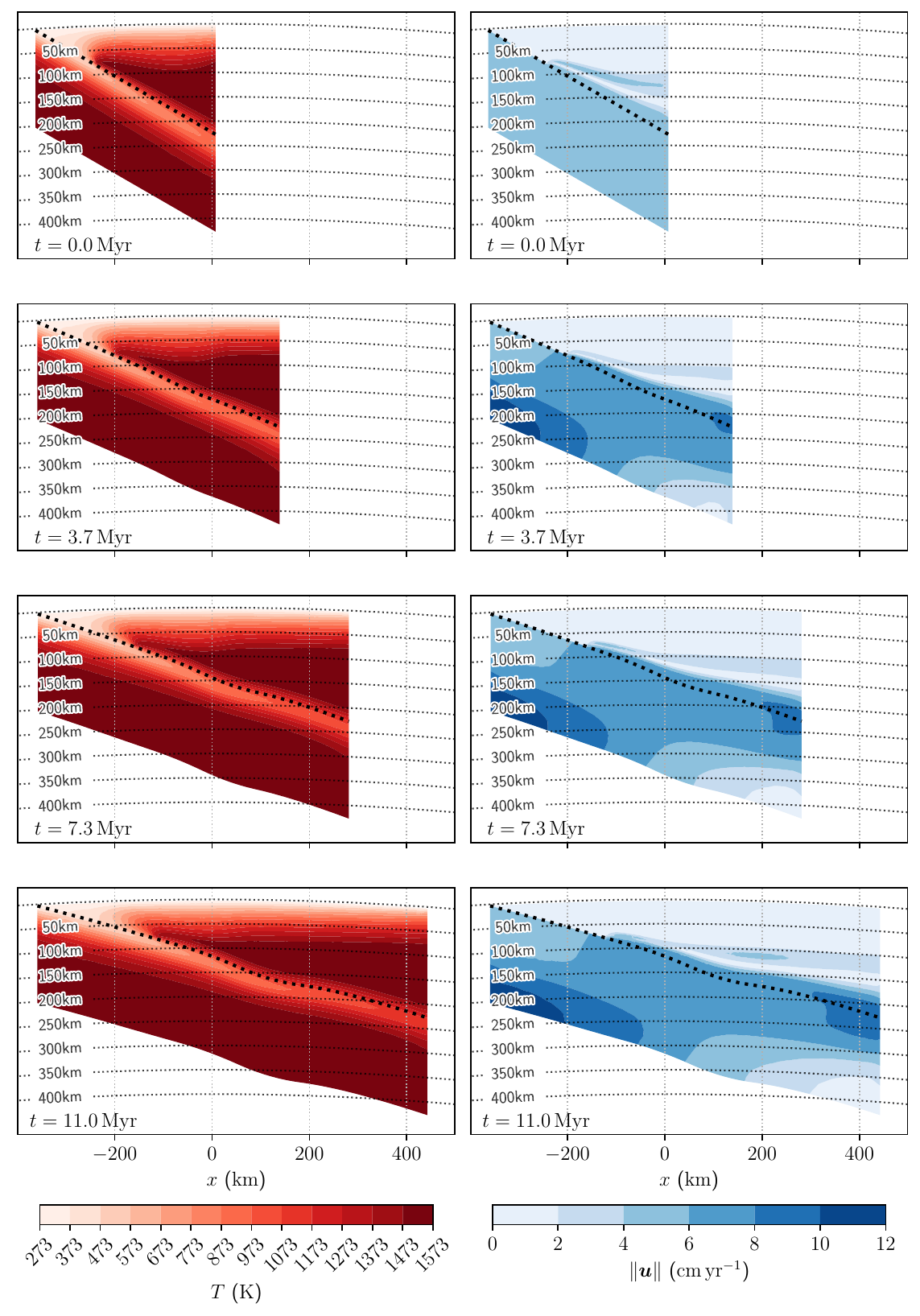}
\caption{Cross-sectional temperature and speeds of the computed $\mathcal{d} = 3$ model.
The data presented is measured at the cross-section of the
full volume parallel to the $y$-axis at location $y=\SI{-200}{\kilo\meter}$ (compare
to $\mathcal{d}=2$ model in \cref{fig:flatslab_temperature_-200}).
The time domain is discretized with \num{100} time steps such that $\Delta t
= \SI{0.11}{\mega\year}$.}
\label{fig:3d_slice_flatslab_temperature_-200}
\end{figure}

\begin{figure}
\centering
\includegraphics[width=.85\linewidth]{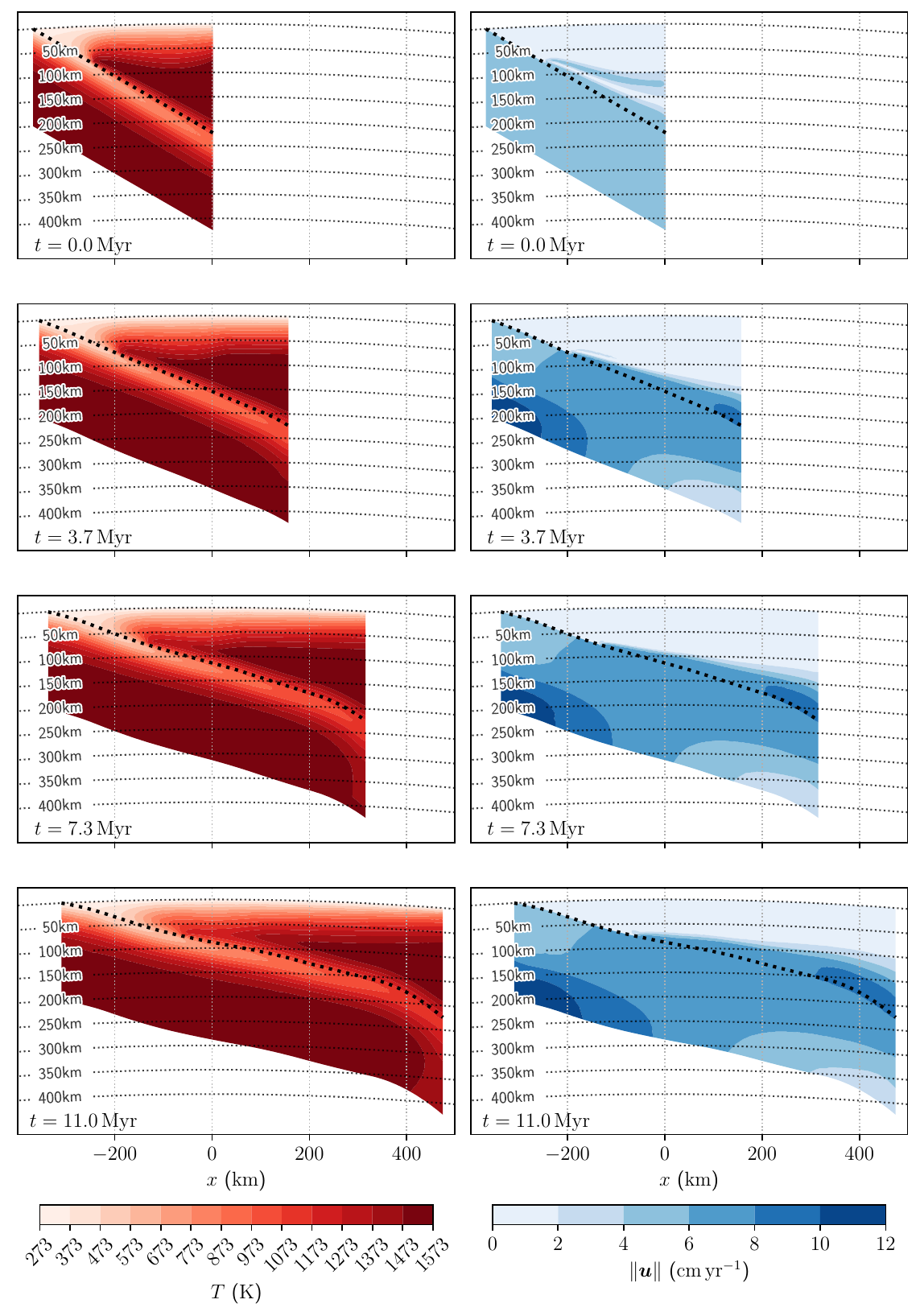}
\caption{As figure \ref{fig:3d_slice_flatslab_temperature_-200} but now at $y=\SI{0}{\kilo\meter}$ (compare
to $\mathcal{d}=2$ model in \cref{fig:flatslab_temperature_0}).}

\label{fig:3d_slice_flatslab_temperature_0}
\end{figure}

\begin{figure}
\centering
\includegraphics[width=.85\linewidth]{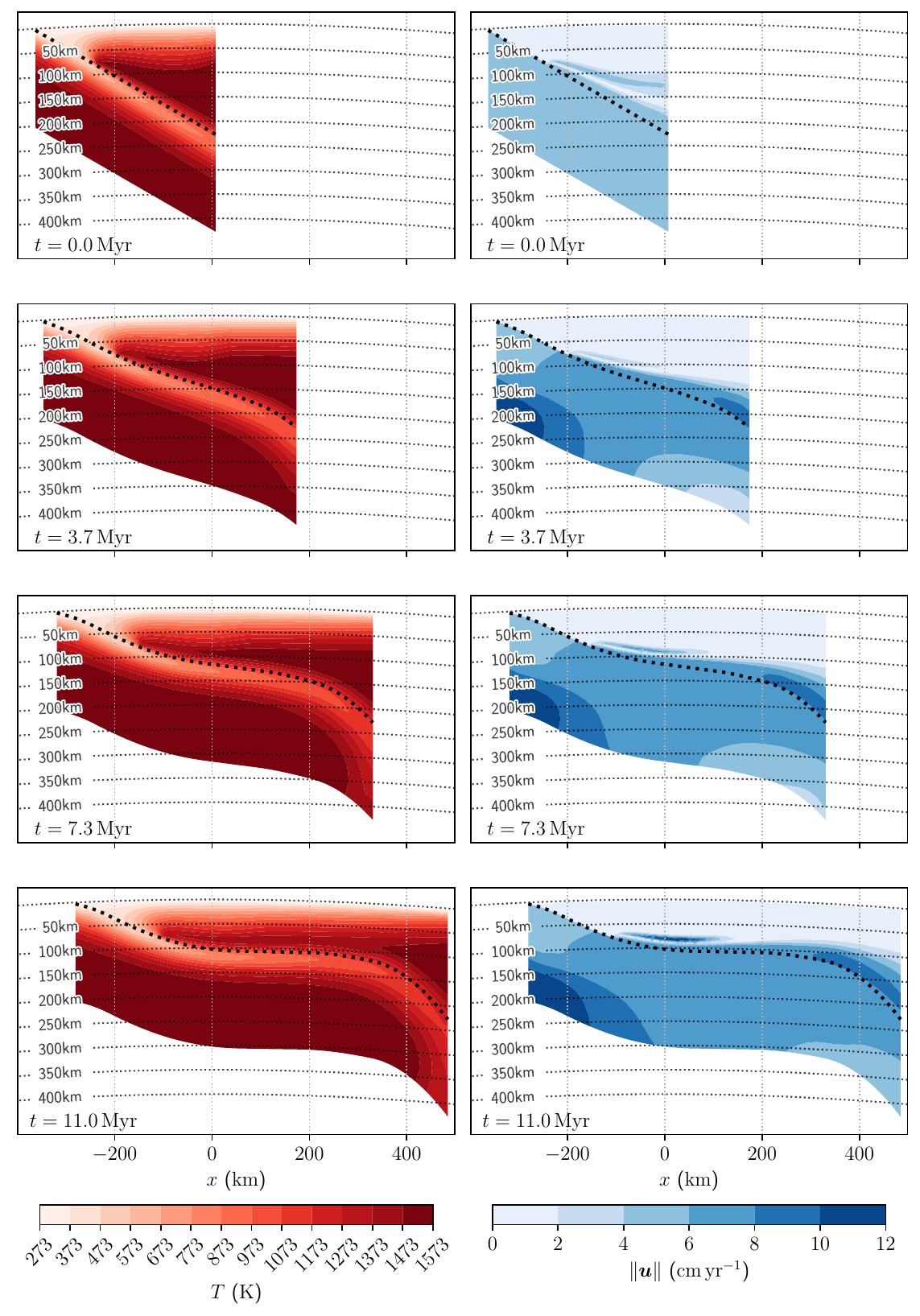}
\caption{As figure \ref{fig:3d_slice_flatslab_temperature_-200} but now at $y=\SI{200}{\kilo\meter}$
(compare to $\mathcal{d}=2$ model in \cref{fig:flatslab_temperature_200}).}
\label{fig:3d_slice_flatslab_temperature_200}
\end{figure}


\begin{figure}
\centering
\includegraphics[width=1.\linewidth]{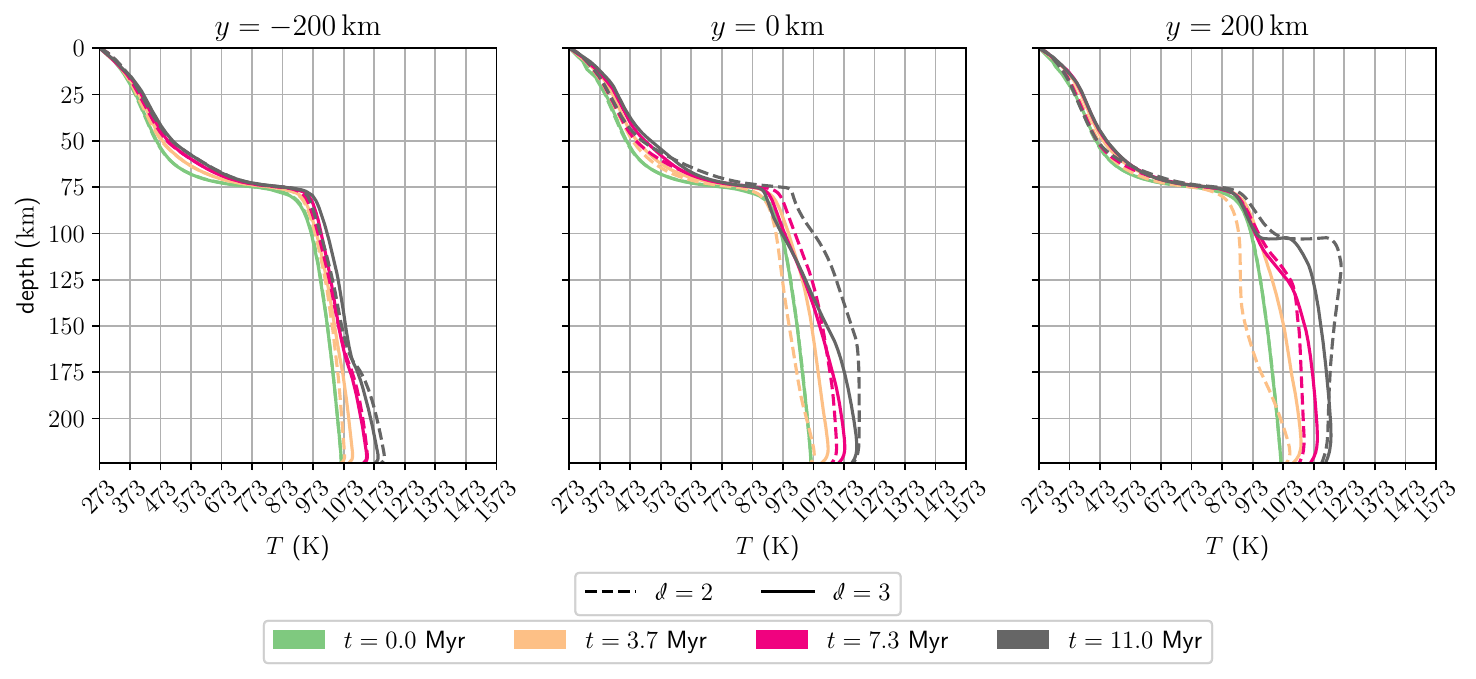}
\caption{Comparison of the slab surface temperatures as a function of depth
as modeled in the $\mathcal{d}=2$ and $\mathcal{d}=3$ cases. In each case, we
plot the surface temperatures as a function of depth along
$\Gamma_\text{slab}$ at the time steps shown in the $\mathcal{d}=2$ models in
\cref{fig:flatslab_temperature_-200,fig:flatslab_temperature_0,fig:flatslab_temperature_200}
and the cross-sections of the $\mathcal{d}=3$ model in
\cref{fig:3d_slice_flatslab_temperature_-200,fig:3d_slice_flatslab_temperature_0,fig:3d_slice_flatslab_temperature_200}.
The difference between the $\mathcal{d}=2$ and $\mathcal{d}=3$ models at time
$t=\SI{0.0}{\mega\year}$ is indistinguishable at this scale.}
\label{fig:flatslab_temperature_2d3d}
\end{figure}

\begin{figure}

\centering
\includegraphics[width=1.\linewidth]{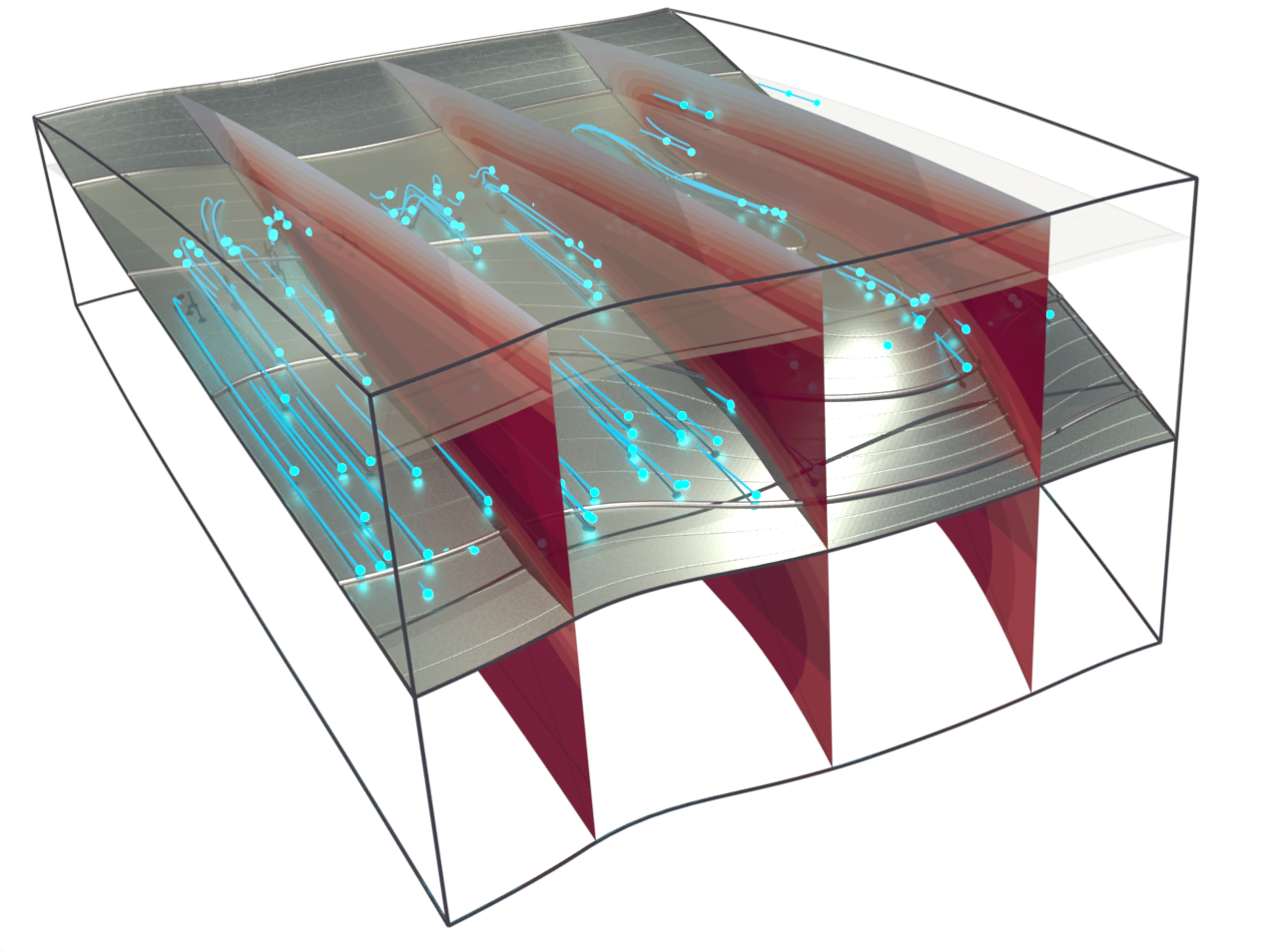}
\caption{Rendering of the $\mathcal{d}=3$ model at time $t=t_\text{slab}$.
Tracers are added where tails show \SI{5.5}{\mega\year} long pathlines. Some
pathlines cross through the slab interface shown; however, this is not an
indication that the tracers have passed through the interface as the time
evolving slab is not shown. There is very minor movement in the tracers
located in the $\Omega_\text{plate}$, their displacement is dictated by the
slab deformation above the plate depth $d_\text{plate}$ highlighted as a
translucent layer. Pathlines close to the slab surface are dominated by the
convergence component of the velocity boundary condition direction
$\hat{\vec{d}}_\text{conv} = \hat{\vec{x}}$.}
\label{fig:flatslab_3draytrace}
\end{figure}





\end{backmatter}
\end{document}